\documentclass[manuscript]{aastex}

\usepackage{graphics,color}
\usepackage{epstopdf}
\definecolor{red}{rgb}{1,0,0}

\shorttitle{Asteroid Rotation Periods From iPTF}
\shortauthors{Chang et al.}
\begin{document}
\title{Large Super-Fast Rotator Hunting Using the Intermediate Palomar Transient Factory}

\author{Chan-Kao Chang\altaffilmark{1}, Hsing-Wen Lin\altaffilmark{1}, Wing-Huen Ip\altaffilmark{1,2}, Thomas A. Prince\altaffilmark{3}, Shrinivas R. Kulkarni\altaffilmark{3}, David Levitan\altaffilmark{3}, Russ Laher\altaffilmark{4}, Jason Surace\altaffilmark{4}}

\altaffiltext{1}{Institute of Astronomy, National Central University, Jhongli, Taiwan}
\altaffiltext{2}{Space Science Institute, Macau University of Science and Technology, Macau}
\altaffiltext{3}{Division of Physics, Mathematics and Astronomy, California Institute of Technology, Pasadena, CA 91125, USA}
\altaffiltext{4}{Spitzer Science Center, California Institute of Technology, M/S 314-6, Pasadena, CA 91125, USA}

\email{rex@astro.ncu.edu.tw}


\begin{abstract}
In order to look for large super-fast rotators, five dedicated surveys covering $\sim 188$~deg$^2$ in the ecliptic plane have been carried out in {\it R}-band with $\sim$ 10 min cadence using the intermediate Palomar Transient Factory in late 2014 and early 2015. Among 1029 reliable rotation periods obtained from the surveys, we discovered one new large super-fast rotator, (40511) 1999 RE88, and other 18 candidates. (40511) 1999 RE88 is an S-type inner main-belt asteroid with a diameter of $D = 1.9 \pm 0.3$ km, which has a rotation period of $P = 1.96 \pm 0.01$ hr and a lightcurve amplitude of $\Delta m \sim 1.0$ mag. To maintain such fast rotation, an internal cohesive strength of $\sim 780~Pa$ is required. Combining all known large super-fast rotators, their cohesive strengths all fall in the range of 100 to 1000 $Pa$ of lunar regolith. However, the number of large super-fast rotators seems to be far less than the whole asteroid population. This might indicate a peculiar asteroid group for them. Although the detection efficiency for a long rotation period is greatly reduced due to our two-day observation time span, the spin-rate distributions of this work show consistent results with \citet{Chang2015} after considering the possible observational bias in our surveys. It shows a number decrease with increase of spin rate for asteroids with diameter of $3 \le D \le 15$ km, and a number drop at spin-rate of $f = 5$ rev/day for asteroids with $D \le 3$ km.
\end{abstract}


\section{Introduction}
It was once a formidable task to collect together a large number of asteroid rotation periods, but it is getting easier. With advances in observational technology, several data sets containing hundreds to thousands of asteroid rotation periods have been acquired through large sky surveys \citep{Masiero2009, Polishook2009, Dermawan2011, Polishook2012, Chang2014a, Chang2015}. Moreover, numerous asteroid rotation periods, obtained from various time-series archived data products \citep[see an example of][]{Waszczak2015}, and single target observations from the Asteroid Light Curve Database \citep[LCDB;][]{Warner2009}\footnote{http://www.minorplanet.info/lightcurvedatabase.html}, also contribute a major portion to this field. Therefore, a more comprehensive understanding of asteroid rotations has emerged and possible applications could be conducted as well \citep[see an example of ][]{Chang2016}.

While the spin-rate distributions obtained from large sky-coverage surveys and archived data show a number decrease with spin rate at frequency $f > 5$ rev/day \citep{Masiero2009, Chang2015, Waszczak2015}, a flat distribution was indicated by the single target observations \citep[][update 2014-04-20]{Pravec2008}. However, the tendency of number decrease still remains in the asteroid spin-rate distributions from the large sky surveys after taking into account possible observational bias \citep{Masiero2009, Chang2015}. In addition, the spin-rate distribution of asteroids with diameter of $D \le 3$ km seems to have a number drop at $f = 5$ rev/day. This could be a result of the
Yarkovsky-O'Keefe-Radzievskii-Paddack effect \citep[YORP;][]{Rubincam2000} that works relatively fast on small asteroids and pushes more of them over the spin-barrier to break up \citep{Chang2015}.

The ``spin-barrier'' at 2.2 hour \citep{Harris1996, Pravec2002} is persistently seen in these data sets. Furthermore, with sufficiently large samples, the C-type asteroids were for the first time found to show a the rotation period limit longer than the S-type asteroids \citep{Chang2015, Waszczak2015}. This is in accordance with the general picture for rubble-pile asteroids that: (a) they cannot rotate exceedingly fast; and (b) those with lower bulk density should have a longer rotation-period limit \citep[$P \sim 3.3 \sqrt{(1 + \Delta m)/\rho}$;][]{Harris1996}. However, the presence of four large (i.e., a few hundreds meter) super-fast rotators (hereafter, SFR), 2001 OE84 \citep{Pravec2002}, 2005 UW163 \citep{Chang2014b}, 1950 DA\citep{Rozitis2014} and 2000 GD65 \citep{Polishook2016}, suggest that internal cohesion might be required to keep them from breaking apart \citep{Holsapple2007, Sanchez2012}. Compared to the majority of large asteroid with known rotation periods, the number of detected large SFR seems to be very small. Therefore, a comprehensive census of the population of large SFR should provide key information on asteroid interior structure.

Therefore, five asteroid rotation-period surveys were carried out to look for large SFRs. We obtained 7984 asteroid lightcurves of $\geq 10$ detections, from which 1029 reliable rotation periods were derived. Among them, we discovered one new large SRF, (40511) 1999 RE88 and other 18 candidates. In this work, the observation information and method of lightcurve extraction are given in Section 2. The rotation-period analysis is described in Section 3. The results and discussion are given in Section 4. A summary and conclusion is presented in Section 5.

\section{Observations and Data Reduction}
To explore the the transient and variable sky synoptically, the PTF/iPTF employs the Palomar 48-inch Oschin Schmidt Telescope equipped with an 11-chip mosaic CCD camera (note that the eleventh chip went out of service in early 2015, and the available chips were therefore ten at that time) to create a field of view of $\sim7.26$ \,deg$^2$ and a pixel scale of 1.01\arcsec~\citep{Law2009, Rau2009}. The available filters include Mould-{\it R} band, with which most exposures were taken, Gunn-{\it g\arcmin}, and two different $H_{\alpha}$ bands. The exposure time is fixed at 60 seconds, which can reach a median limiting magnitude of $R \sim $21 mag at the $5\sigma$ level \citep{Law2010}.

All PTF/iPTF exposures are processed by the IPAC-PTF photometric pipeline \citep{Grillmair2010, Laher2014}, and the absolute magnitude, calibrated against Sloan Digital Sky Survey fields \citep[hereafter, SDSS;][]{York2000}, can routinely reach a precision of $\sim0.02$~mag on photometric nights \citep{Ofek2012a, Ofek2012b}. Since the magnitude calibration is based on a per-night, per-filter, per-chip basis, small photometric zero-point variations are present in catalogs for different nights, fields, filters and chips.

In order to look for large SFRs, we conducted five asteroid rotation period surveys during October 29-31 and November 10-13 of 2014, and January 18-19, February 20-21 and February 25-26 of 2015. Each survey continuously scanned six consecutive
PTF fields over the ecliptic plane in the $R$ band with a cadence of 10 min. While the first two surveys in late 2014 were observed in three straight nights, the last three were only observed in two adjacent nights. However, there were only several exposures in the first and last nights for November 2014 observation due to bad weather condition. We finally ended up with a total sky coverage of $\sim 188$~deg$^2$. The observational metadata are listed in Table~\ref{obs_log_01}.

To extract the lightcurves of known asteroids, we removed the stationary sources from the catalogs and then matched the detections against the ephemerides obtained from the {\it JPL/HORIZONS} system with a search radius of 2\arcsec. We also excluded the detections flagged as a defect by the IPAC-PTF photometric pipeline from the lightcurves. Finally, there were 7914 asteroid lightcurves with number of detections of $\geq 10$ (hereafter, PTF-detected asteroids) for the rotation-period analysis described in the next section.

\section{Rotation-Period Analysis}\label{period_analysis}
All the measurements in the lightcurves were corrected for light-travel time and reduced to both heliocentric, $r$, and geocentric, $\triangle$, distances at 1~AU. Since the changes of phase angles is small during our observation time span, we simply estimated the absolute magnitude by applying a fixed $G_R$ slope of 0.15 in the $H$--$G$ system \citep{Bowell1989}. Then, we followed the traditional second-order Fourier series method to derive the rotation period \citep{Harris1989}:
\begin{equation}\label{FTeq}
  M_{i,j} = \sum_{k=1,2}^{N_k} B_k\sin\left[\frac{2\pi k}{P} (t_j-t_0)\right] + C_k\cos\left[\frac{2\pi k}{P} (t_j-t_0)\right] + Z_i,
\end{equation}
where $M_{i,j}$ are the $R$-band reduced magnitudes measured at the light-travel time corrected epoch, $t_j$; $B_k$ and $C_k$ are the Fourier coefficients; $P$ is the rotation period; and $t_0$ is an arbitrary epoch. The constant values, $Z_i$, are introduced to correct the small aforementioned photometric zero-point variations. The least-squares minimization was applied to Eq.~(\ref{FTeq}) to obtain the other free parameters for each given $P$. The spin-rate, $f$, was searched from 0.25 to 25~rev/day with a step size of 0.025~rev/day.

A quality code ($U$) was then manually assigned to each folded lightcurve by visual inspection, where: `3'~means highly reliable; `2'~means some ambiguity; `1'~means possible, but may be wrong \citep{Warner2009}. Moreover, when no acceptable solution can be found for a lightcurve, it was assigned $U = 0$. The uncertainty of the derived rotation period was estimated from periods having $\chi^2$ smaller than $\chi_{best}^2+\triangle\chi^2$, where $\chi_{best}^2$ is the $\chi^2$ of the derived period and $\triangle\chi^2$ is calculated from the inverse $\chi^2$ distribution, assuming $1 + 2N_k + N_i$~degrees of freedom. We adopted the peak-to-peak amplitude after rejecting the upper and lower 5\% of data points to avoid outliers, which are probably contaminated by nearby bright stars or unfiltered artifacts during the lightcurve extraction.

Moreover, we adopted $WISE$/$NEOWISE$ diameter estimation, if available, for PTF-detected asteroids \citep{Grav2011, Mainzer2011, Masiero2011}. Otherwise, the diameter was estimated using
\begin{equation}\label{dia_eq}
  D = {1130 \over \sqrt{p_R}} 10^{-H_R/5},
\end{equation}
where $H_R$ is the $R$ band absolute magnitude, $D$ is the diameter in~km, $p_R$ is the $R$ band geometric albedo, and 1130 is the conversion constant adopted from \citet{Jweitt2013}. Three empirical albedo values, $p_R =0.20$, 0.08 and 0.04, were assumed for asteroids in the inner ($2.1 < a < 2.5 AU$), mid ($2.5 < a < 2.8$ AU) and outer ($a > 2.8$ AU) main belts, respectively \citep{Tedesco2005}.

\section{Results and Discussion}
\subsection{Derived Rotation Periods}\label{discuss_p}
\begin{figure}
\plotone{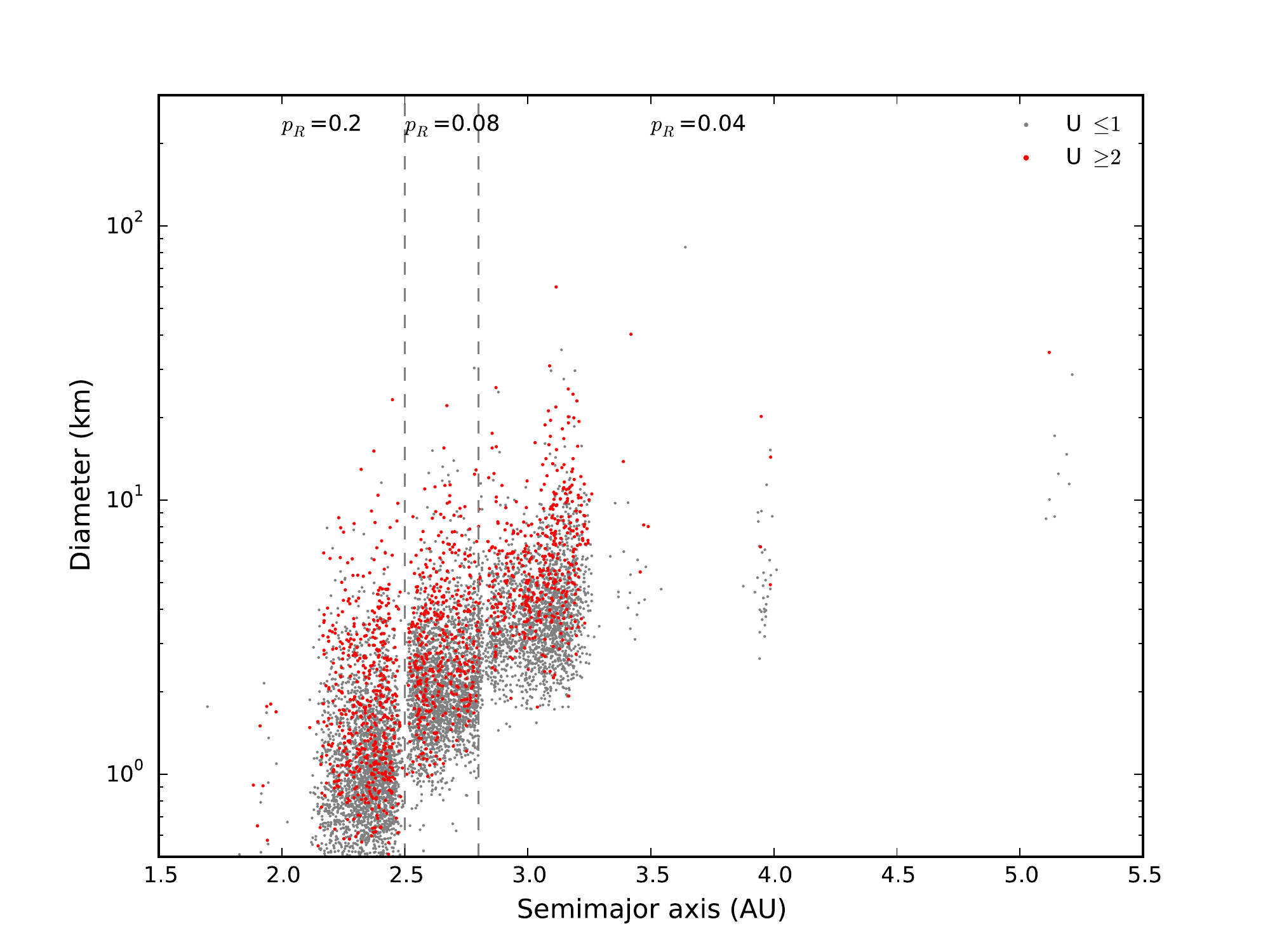}
\caption{Diameters vs.\ semi-major axes for the PTF-U2s (red) and the PTF-detected asteroids (gray).
The dashed lines show the divisions of empirical geometric albedo ($p_R$) for asteroids located at
different regions of the semi-major axis.}
\label{a_d}
\end{figure}

\begin{figure}
\plotone{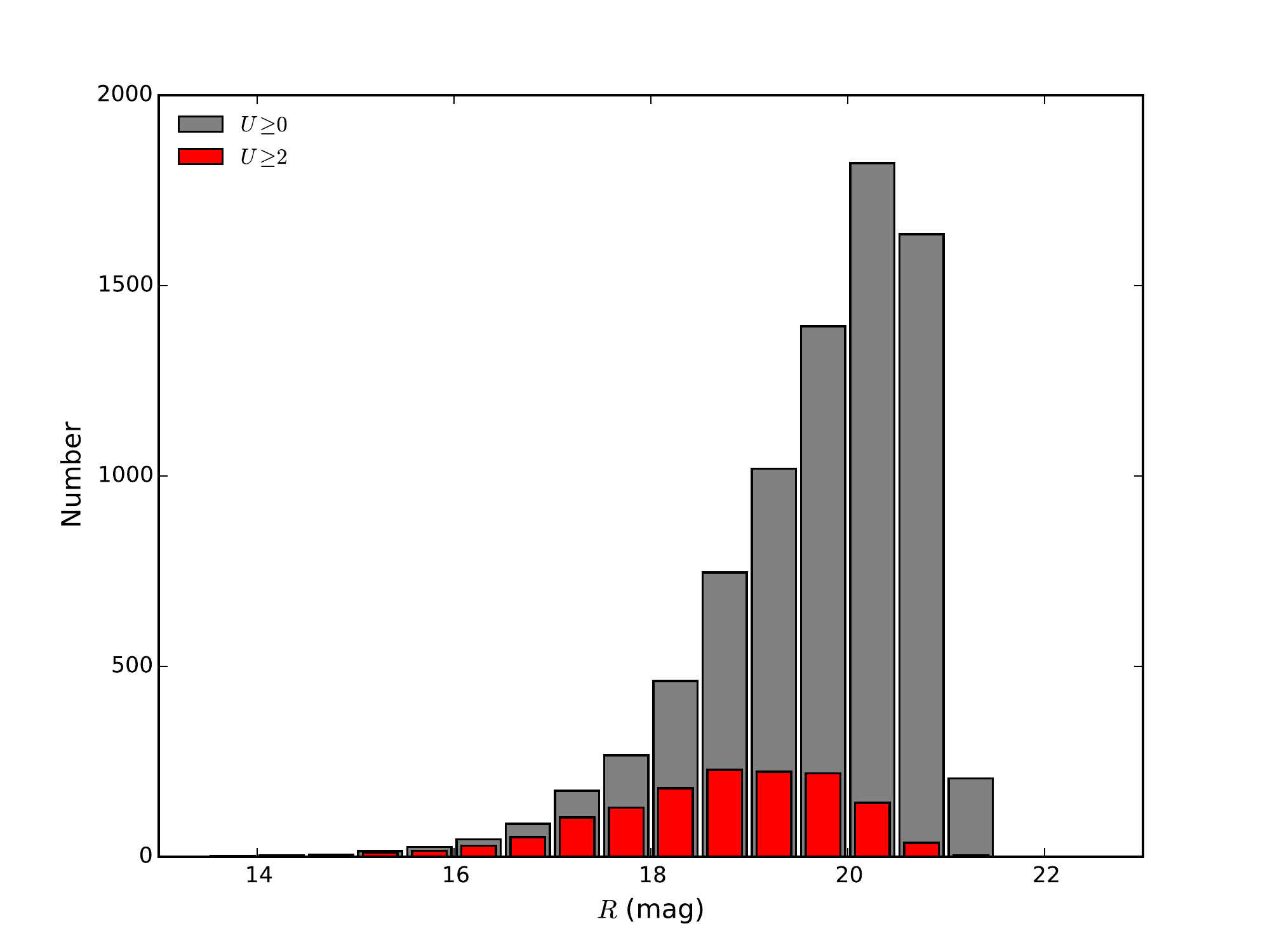}
\caption{Magnitude distributions of the PTF-detected asteroids (gray) and the PTF-U2s (red) of this works. }
\label{mag_his}
\end{figure}

\begin{figure}
\includegraphics[angle=0,scale=.7]{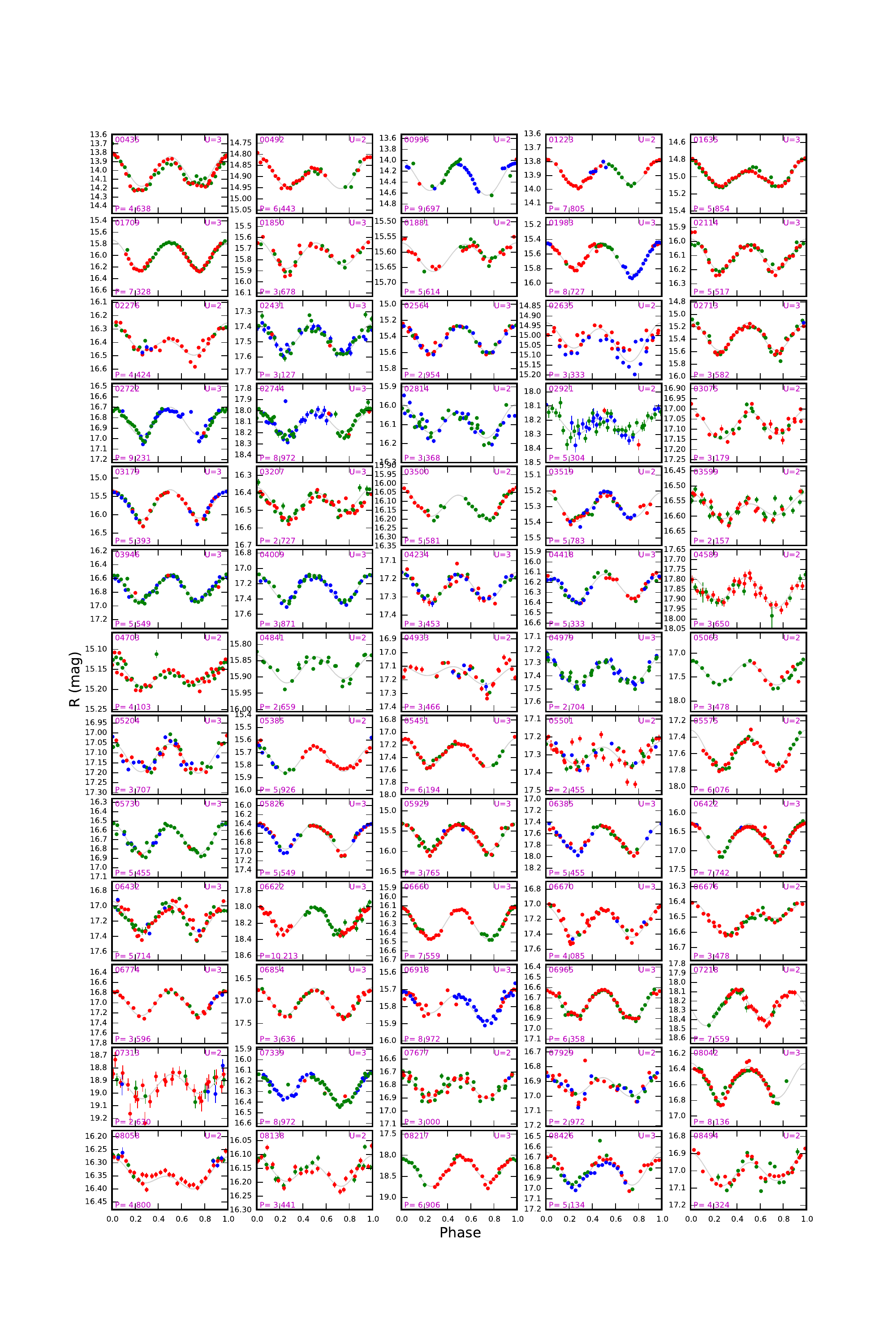}
\caption{Set of 65 folded lightcurves for the PTF-U2s. Colors represent observations taken in different nights. The asteroid designation is given on each plot along with its derived rotation period $P$ in hours and quality code $U$.}
\label{lightcurve00}
\end{figure}

\begin{figure}
\includegraphics[angle=0,scale=.7]{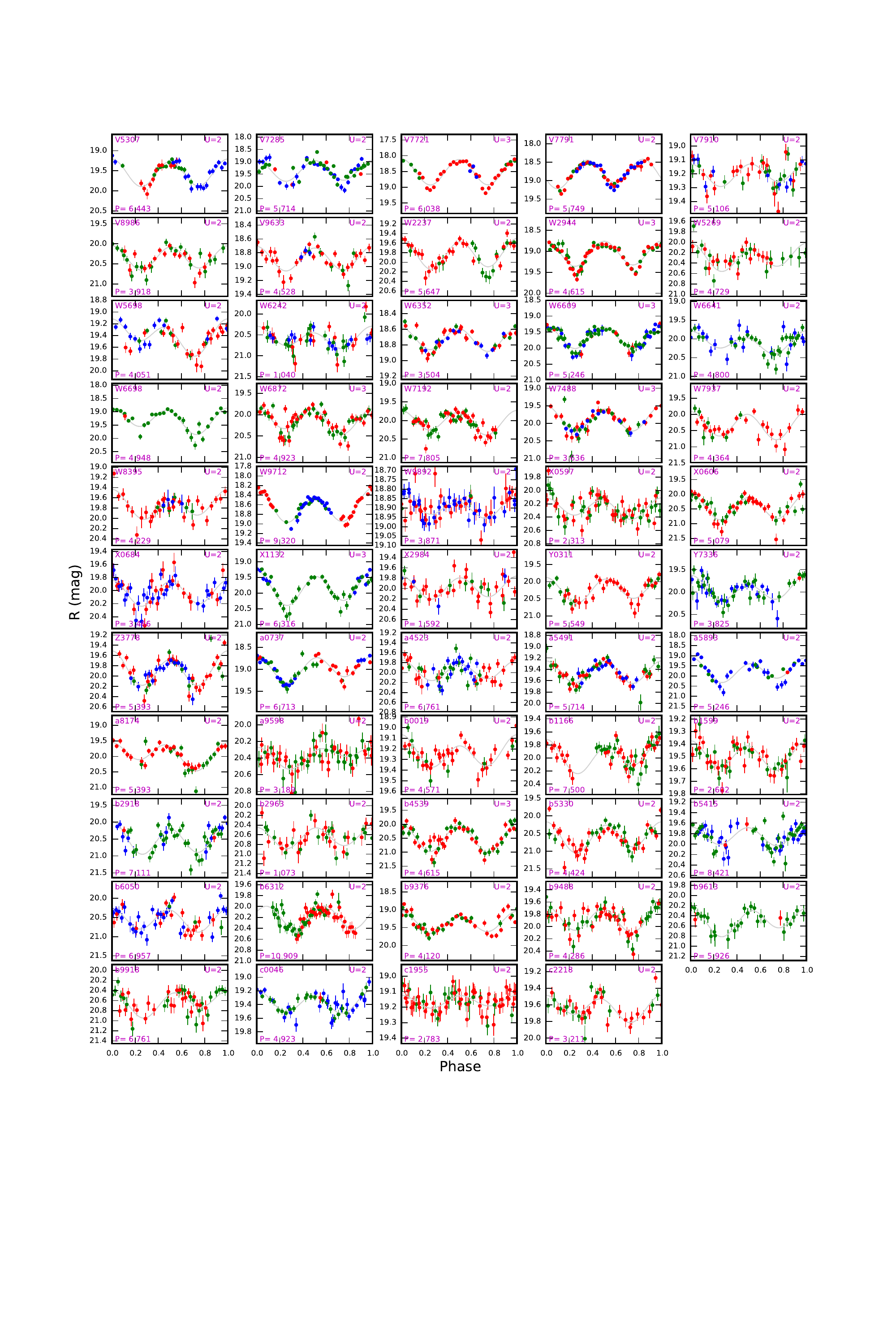}
\caption{Same as Fig.~\ref{lightcurve00} for other 54 PTF-U2s.}
\label{lightcurve15}
\end{figure}

\begin{figure}
\includegraphics[angle=0,scale=.7]{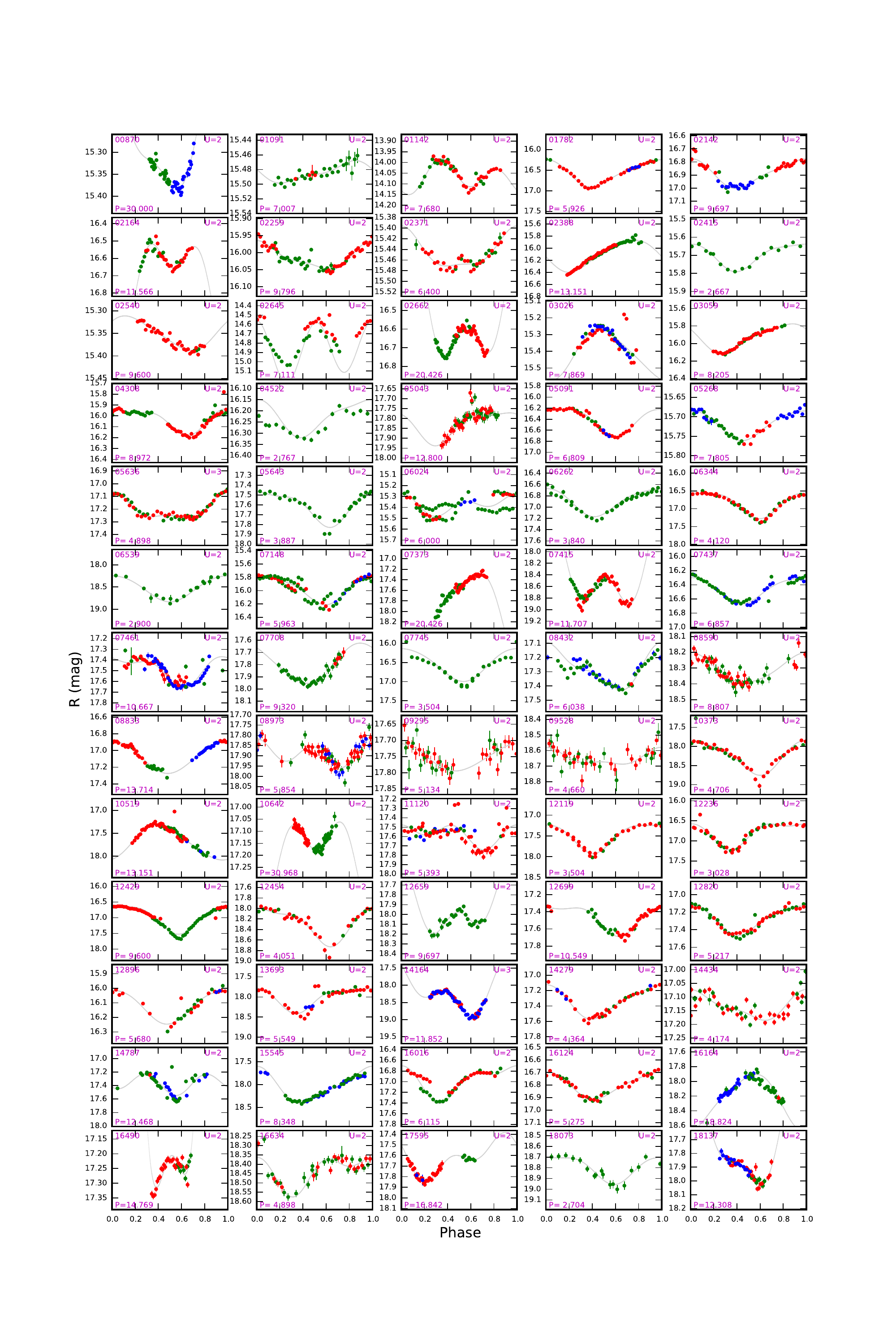}
\caption{Same as Fig.~\ref{lightcurve00} for 65 PTF-Ps.}
\label{lightcurve_p_00}
\end{figure}

\begin{figure}
\includegraphics[angle=0,scale=.7]{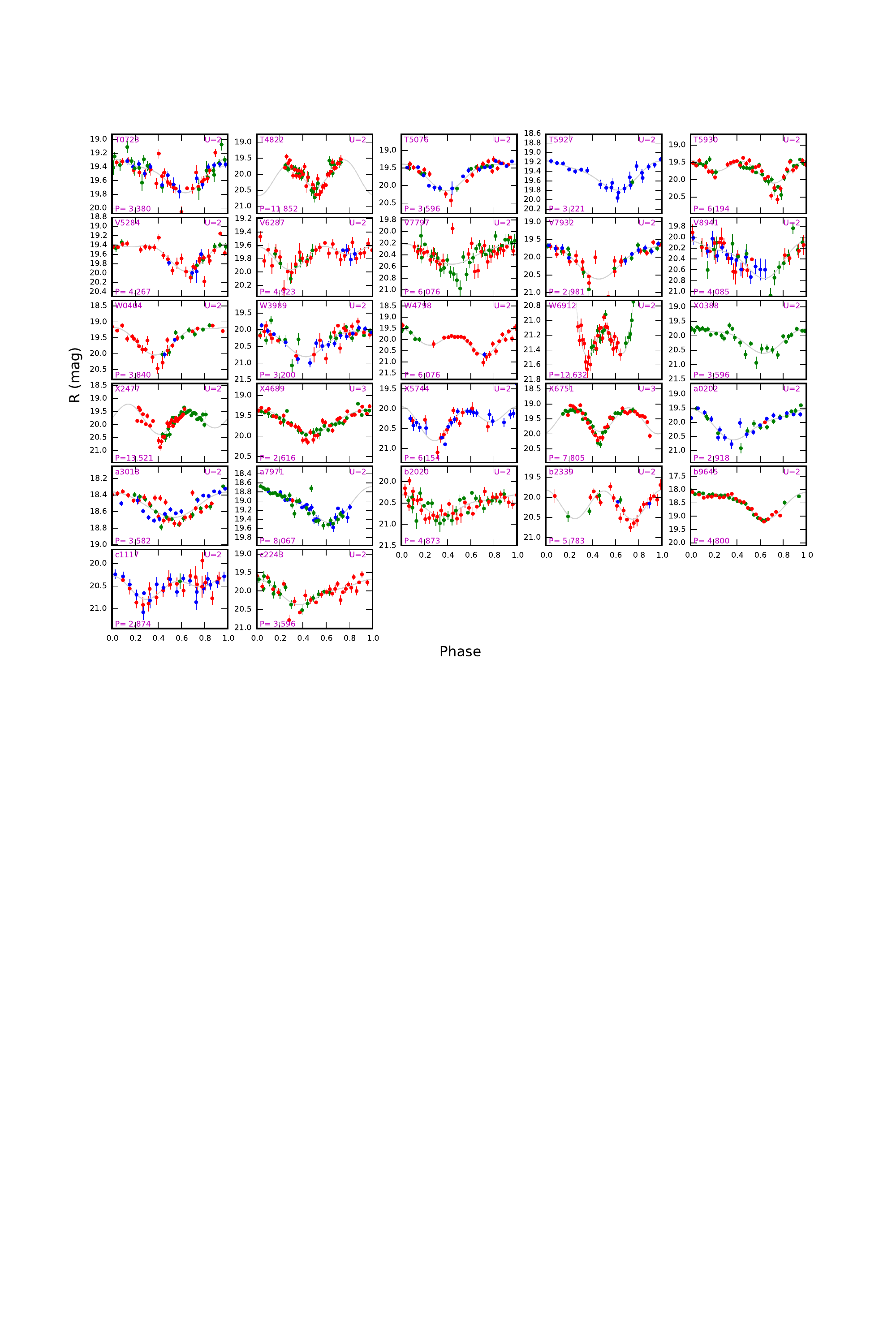}
\caption{Same as Fig.~\ref{lightcurve00} for other 27 PTF-Ps.}
\label{lightcurve_p_05}
\end{figure}

We obtained 1029 reliable (i.e., $U \ge 2$) rotation periods (hereafter, PTF-U2s). Their rotation periods, orbital elements, and observational conditions are summarized in Table~\ref{table_p}, and their folded lightcurves are given in Figs.~\ref{lightcurve00}-\ref{lightcurve15}. Moreover, we also obtained 352 asteroids whose folded lightcurves show a clear trend, but do not fully cover one revolution (hereafter, PTF-Ps). Most of the PTF-Ps seem to have relatively long rotation periods (i.e., $f < 2$ rev/day) that cannot be recovered by our short observation time-span. Therefore, their derived rotation periods can be treated as lower limits. These asteroids are summarized in Table~\ref{table_p_part} and their folded lightcurves are given in Fig.~\ref{lightcurve_p_00}-\ref{lightcurve_p_05}. Because of the survey area and the limiting magnitude, the majority of our samples are main-belt asteroids, as shown by the distribution of diameters vs. semi-major axes in Fig.~\ref{a_d}. As expected, the chance of recovering the rotation period is better for brighter objects as seen in Fig.~\ref{mag_his}, which shows the overall magnitude distribution of the PTF-U2s and PTF-detected asteroids.

\begin{figure}
\plotone{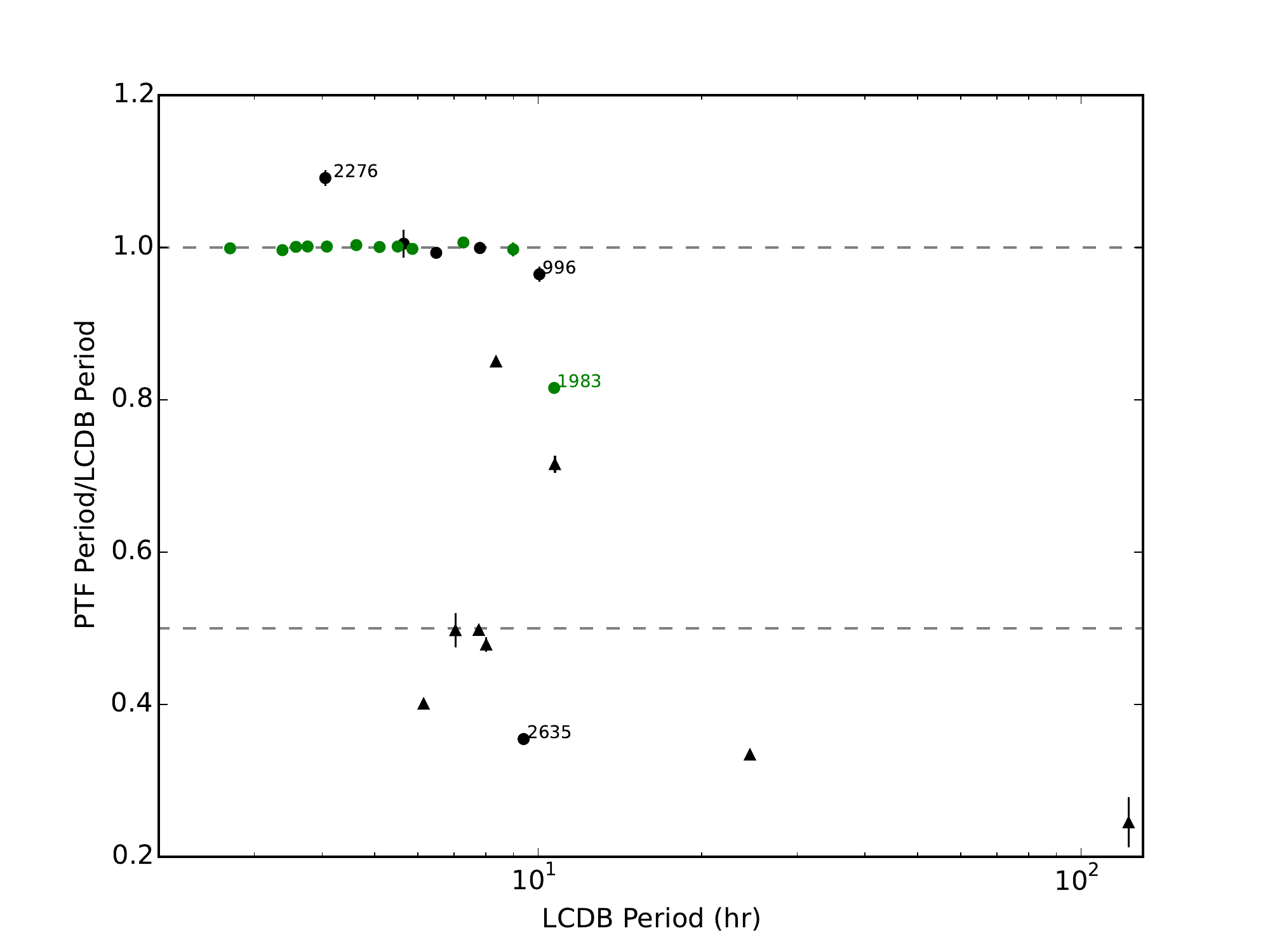}
\caption{Comparison of 26 rotation periods of the PTF-U2s and the LCDB asteroids of $U = 3$. Filled
circles and filled triangles correspond to the PTF-U2s and the PTF-Ps, respectively. Green and black indicate $U$ of the PTF-U2 is equal or worse than the matching LCDB object, respectively.}
\label{diff_p}
\end{figure}

To examine our derived rotation periods, we compare our results with the $U = 3$ asteroids in the LCDB. There are 26 overlapping objects, and the comparison is shown in Fig.~\ref{diff_p}. In general, (a) 14 out of 17 derived rotation periods of the PTF-U2s are in good agreement with the LCDB values (i.e., the difference is $<$ 3\%); (b) since the PTF-Ps have relatively large uncertainty in their derived rotation periods due to the incomplete lightcurve coverage on their full revolutions, all the PTF-Ps in Fig.~\ref{diff_p} show certain degree of difference to the LCDB values; and (c) only two PTF-U2s, asteroid 996 and 2267, show minor differences (i.e., $\lesssim 10\%$) and one PTF-U2, asteroid 2635, has a large difference with respect to the LCDB value. We discuss these three objects below. The $U$ codes of these three objects in our results were assigned as 2, which means that these three objects have relatively large uncertainties in our results. In fact, our derived rotation period for asteroid 996 (i.e., 9.70 hr for the  PTF-U2s and 10.05 hr in the LCDB) and asteroid 2276 (i.e., 4.42 hr for the PTF-U2s and 4.05 hr in the LCDB) are in very good agreement with the LCDB values. However, our two-day observation time-span was just long enough to merely cover the whole revolution of asteroid 996 and consequently leads to a shorter period. When we re-examined the periodogram for asteroid 2276, we found that its PTF lightcurve could be folded equally well on the periods of 4 and 4.8 hr besides the best-fit 4.42 hr. The preference of 4.42 hr is due to the resulting less dense lightcurve. When the observations were taken, asteroid 2635 happened to pass by a bright neighboring star and moved to the chip boundary. Consequently, most data points for asteroid 2635 were contaminated and are relatively unreliable. Moreover, it has a small lightcurve amplitude of 0.1 mag \citep{Mazzone2012}, we were therefore unable to identify its true rotation period. Overall, the derived rotation periods of the PTF-U2s are reliable enough to yield the statistics on the asteroid spin rate.

\subsection{Spin-Rate Limit and Large Super-Fast Rotators}
\begin{figure}
\plotone{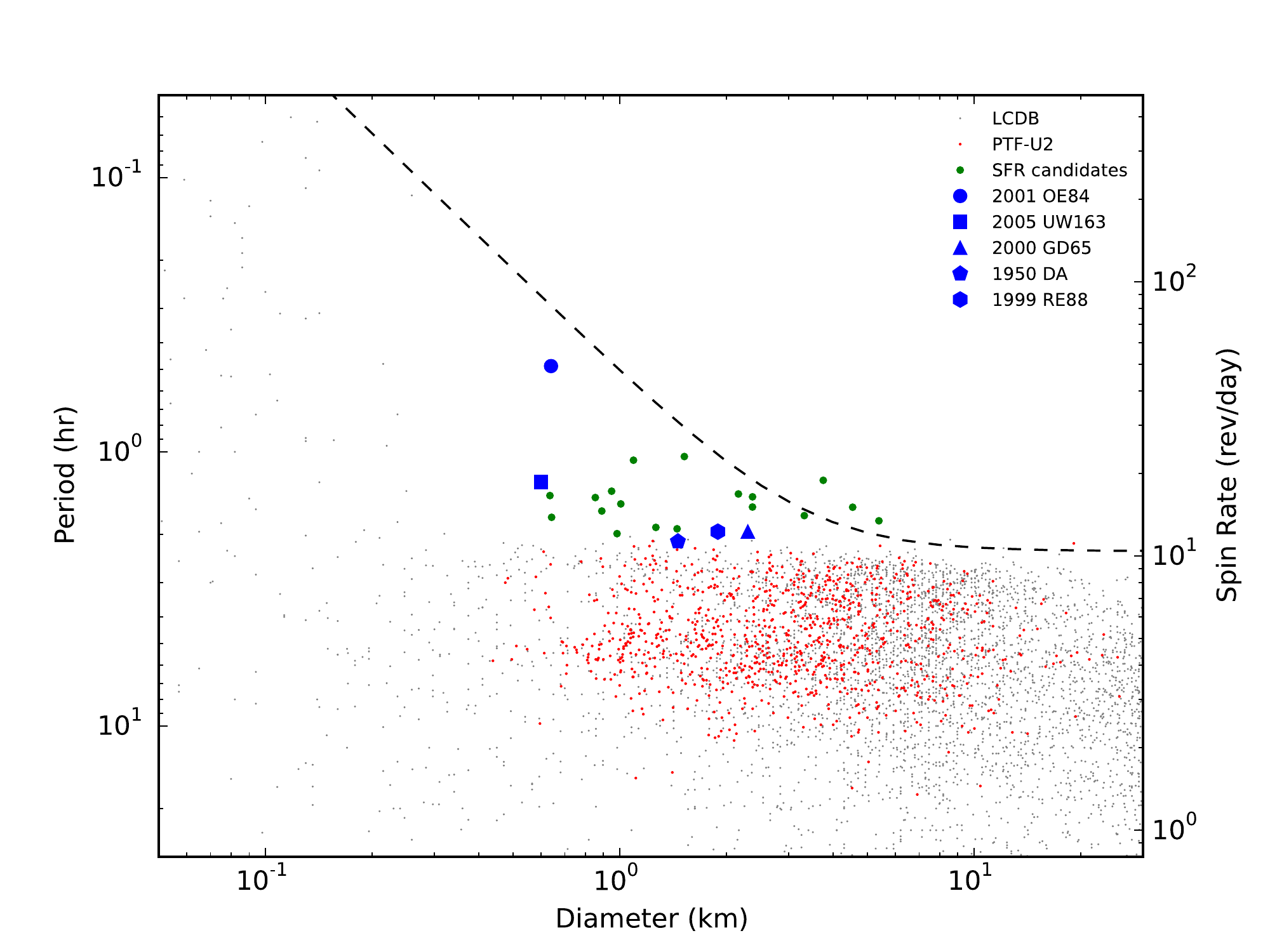}
\caption{Asteroid rotation period vs.\ diameter. The red and gray filled circles are the PTF-U2s and the LCDB objects
of $U \geq 2$, respectively. The reported large SFRs are shown with larger blue filled symbols and the newly discovered large SFR, (40511) 1999 RE88, is represented by the blue filled hexagon. The green filled circles are the SFR candidates from this work. The dashed line is the predicted spin-barrier adopted from \citet{Holsapple2007}. Note that the uncertainties in diameter estimation using Eq.~\ref{dia_eq} for 18 SFR candidates are $\sim 10\%$ according to the uncertainties in their absolute magnitude $H$.}
\label{dia_per}
\end{figure}

\begin{figure}
\includegraphics[angle=0,scale=.7]{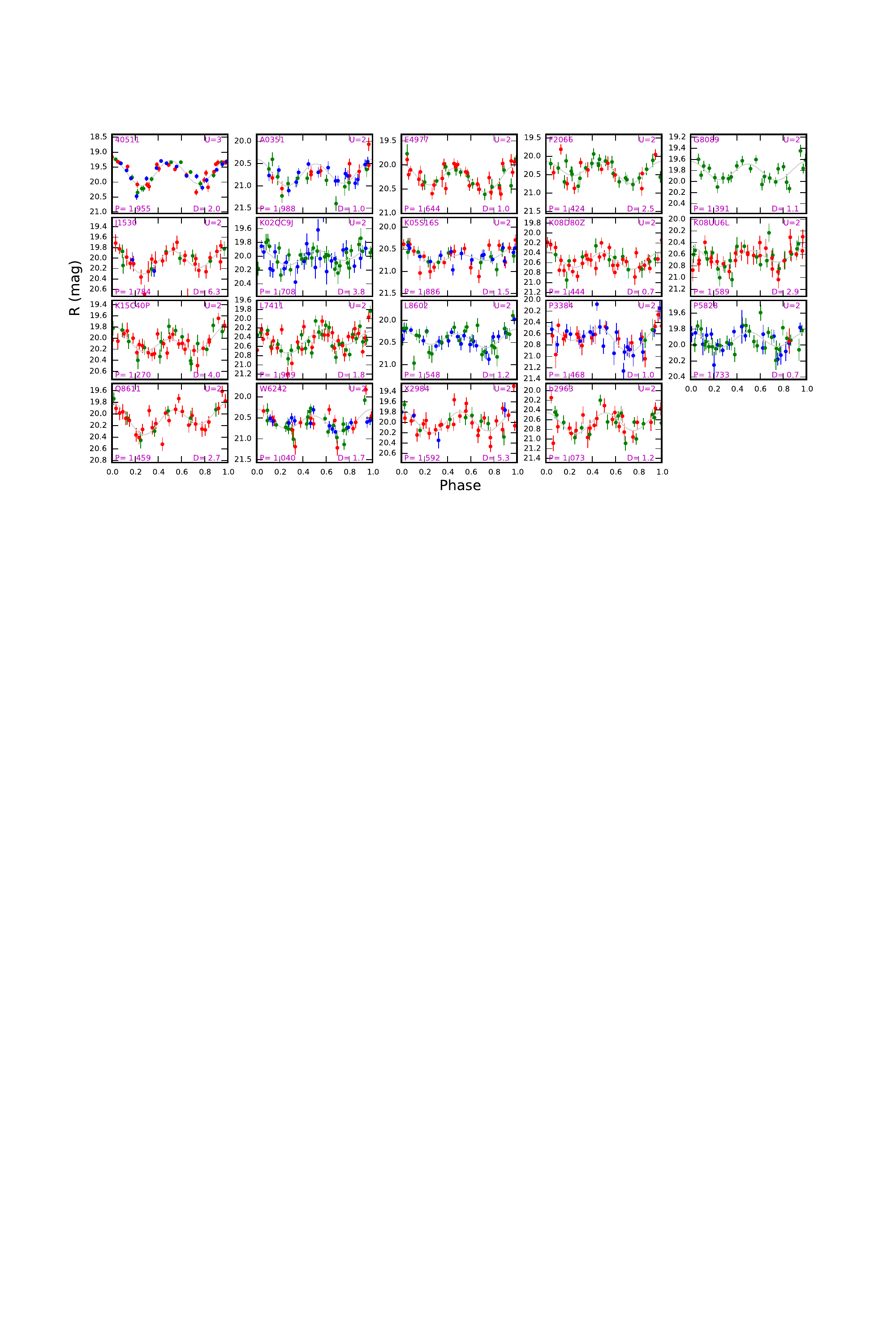}
\caption{The 19 folded lightcurves for (40511) 1999 RE88, and other 18 candidates. Colors represent
observations taken in different nights. The asteroid designation is given on each plot along with its derived rotation
period $P$ in hours and quality code $U$.}
\label{lightcurveSFR}
\end{figure}

\begin{figure}
\plotone{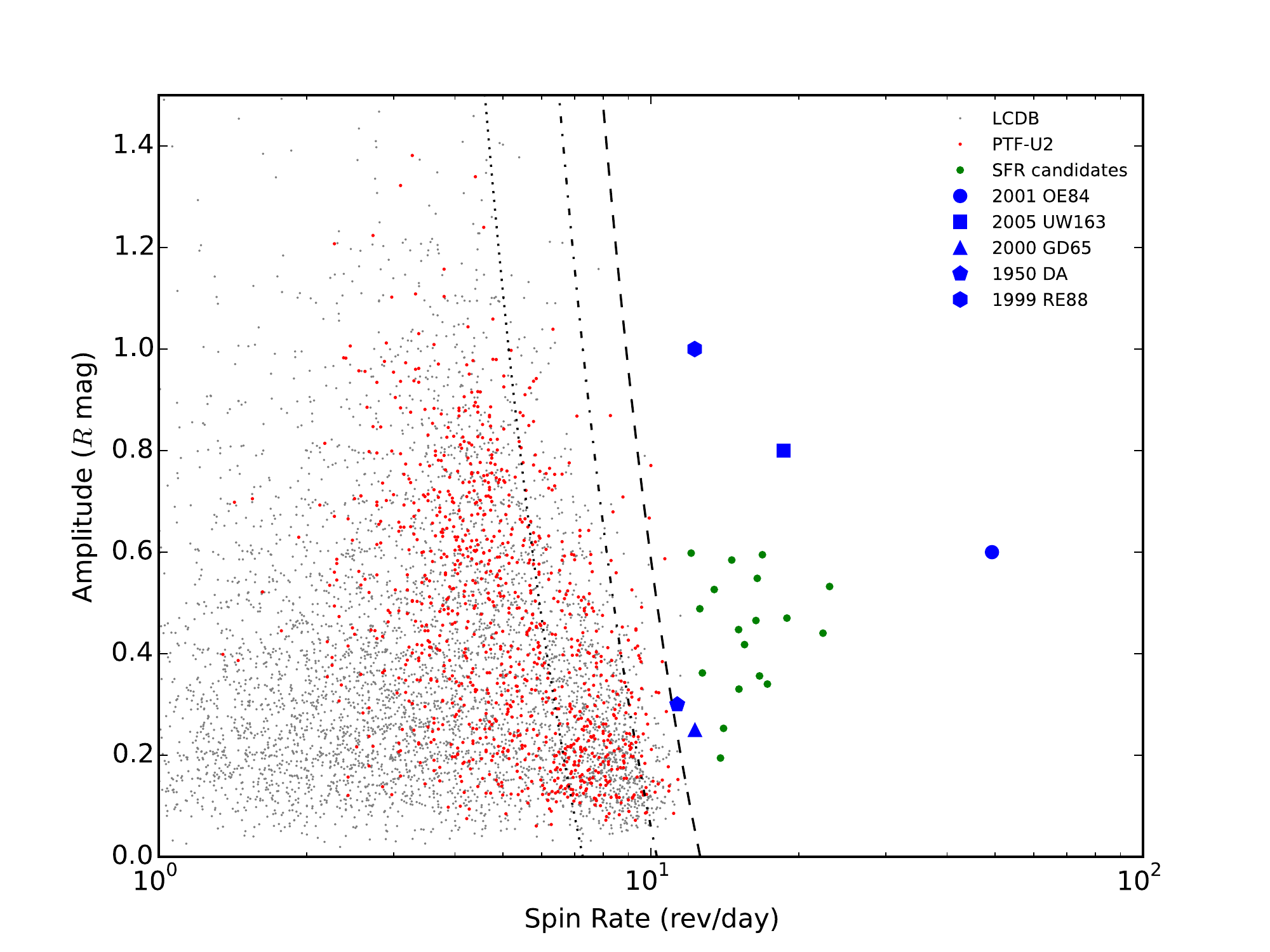}
\caption{Lightcurve amplitude vs.\ spin rate. The symbols are the same with Fig.~\ref{dia_per}. The dashed, dot-dashed and dotted lines represent the spin-rate limits for rubble-pile asteroids with bulk densities of $\rho =$ 3, 2, and 1~g/cm$^3$, respectively, according to $P \sim 3.3 \sqrt{(1 + \Delta m)/\rho}$ \citep{Pravec2000}. Note that the asteroids of $D < 0.2$ km are not included in this plot.}
\label{spin_amp}
\end{figure}

To investigate the spin-rate limit at 2.2 hr, we plot the diameters vs. rotation periods of the PTF-U2s along with that of the asteroids of $U = 3$ in the LCDB. Fig.~\ref{dia_per} shows that most PTF-U2s are still below the 2.2 hr limit, which is in accordance with the rubble-pile structure. However, 19 PTF-U2s with diameters ranging from several hundred meters to several kilometers locate above the limit. We nominate these objects as large SFR candidates and list them separately in Table~\ref{table_sfr}. Their folded lightcurves are given in Fig.~\ref{lightcurveSFR} and all show a clear trend. Asteroids with diameter of $D > 150$ m are believed to be rubble-pile due to their complex collision history \citep{Pravec2002}. These large SFR candidates are of particular interest to the understanding of asteroid interior structure. When $P \sim 3.3 \sqrt{(1+\Delta m)/\rho}$ is applied to the PTF-U2s, the results suggest that these large SFR candidates have a bulk density of $\rho > 3$~g/cm$^3$ as shown in the plot of spin rate vs. amplitude in Fig.~\ref{spin_amp}. Such high bulk density is very unusual and it is therefore believed that internal cohesion might be present in asteroids \citep{Holsapple2007}. However, the large SFRs, including these candidates, seem to comprise far less than the whole population of asteroid. This indicates that the large SFRs might be a special group aside from the ``average'' asteroids, which perhaps possess a different evolutionary history or mechanism to survive under their super-fast rotations.

\subsubsection{The Large Super-Fast Rotators: (40511) 1999 RE88}
\begin{figure}
\plotone{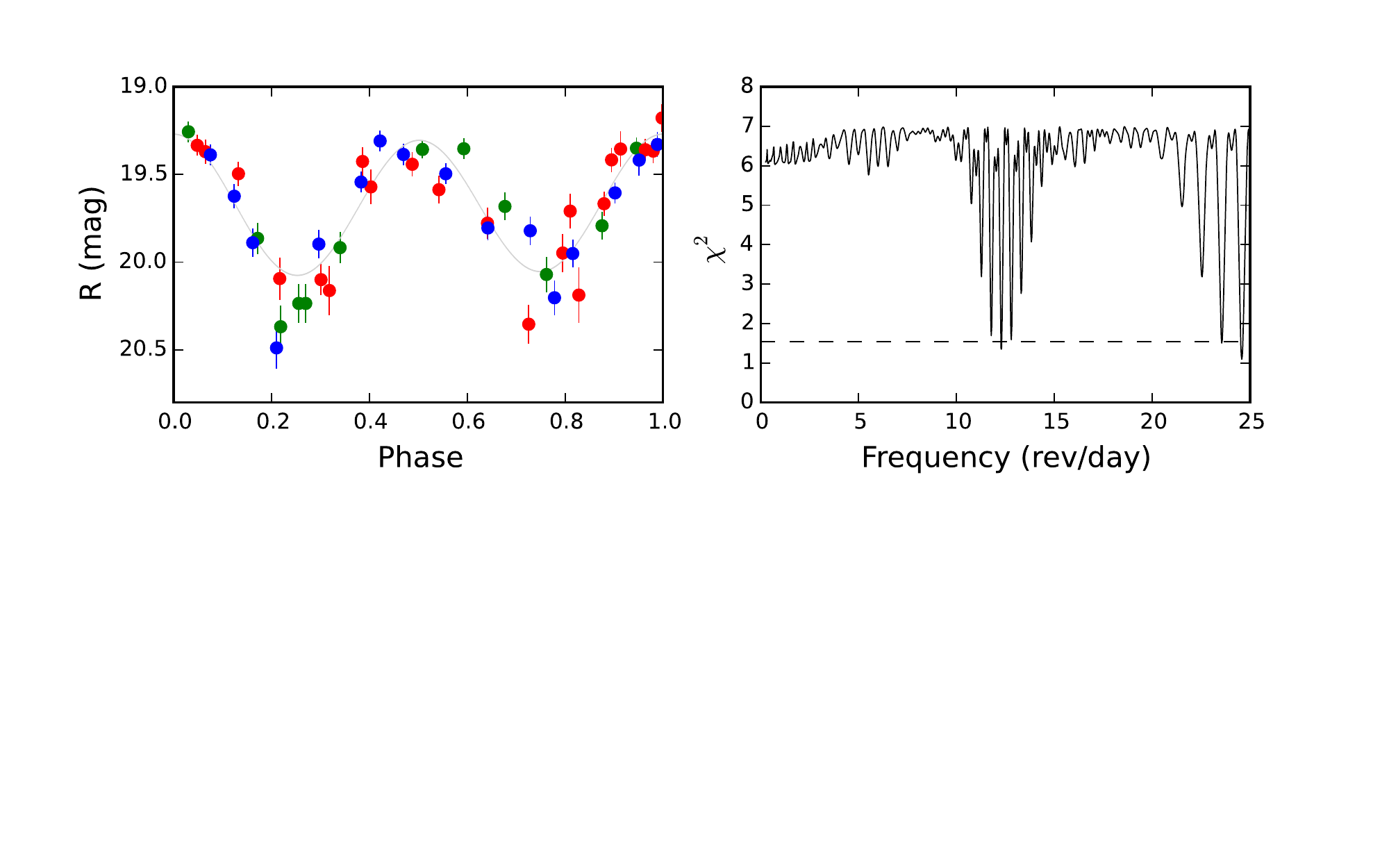}
\caption{The folded lightcurve (left) and the periodogram (right) for the SFR, (40511) 1999 RE 88. Colors in the lightcurve represent observations taken in different nights. The dashed line in the periodograms indicate the uncertainties of the derived rotation periods. Note that an octahedron shape would have maximum amplitude of $< 0.4$ mag \citep{Harris2014}, and therefore, the $\sim 1.0$ mag amplitude of (40511) 1999 RE 88 can rule out the possibility of being at the double spin-rate $f = 24.55$ rev/day.}
\label{lc_40511}
\end{figure}

Among the SFR candidates, the asteroid (40511) 1999 RE88 demonstrates a very clear folded lightcurve on the best-fit period of $1.96 \pm 0.01$ hr (see left panel in Fig.~\ref{lc_40511}). When inspecting its periodogram, 1999 RE88 shows a simple profile with a very significant dip of $\chi^2$ at the best-fit frequency and a relatively high value of mean $\chi^2$ (see right panel Fig.~\ref{lc_40511}). This is very similar to the periodograms of other asteroids with $U = 3$ in the PTF-U2s. The $WISE$/$NEOWISE$ measurement gives it a diameter of $1.9 \pm 0.3$ km. Therefore, we identify (40511) 1999 RE88 as a newly discovered SFR. According to the SDSS color (i.e., $a^* = 0.12\pm0.03$ and $i-z = 0.64 \pm 0.11$) and the $WISE$/$NEOWISE$ albedo (i.e., $p_V = 0.18 \pm 0.04$ and $P_{IR} = 0.27 \pm 0.06$), it suggests that 1999 RE88 is a S-type inner main-belt asteroid. The folded lightcurve amplitude of $\sim 1$ mag rules out the possibility of an octahedron shape for 1999 RE88 \citep{Harris2014}. An asteroid with diameter of 1999 RE88 is very unlikely to be monolithic due to its complex collision history (i.e., $\sim 10^4$ impacts within $10^9$ years) as shown by \citet{Polishook2016}. To calculate the internal cohesion that prevents 1999 RE88 from breaking apart in such fast rotation, we apply the Drucker-Prager yield criterion \citep{Holsapple2007, Rozitis2014, Polishook2016}:
\begin{equation}
  {1 \over 6}[(\overline{\sigma}_x-\overline{\sigma}_y)^2 + (\overline{\sigma}_y-\overline{\sigma}_z)^2 + (\overline{\sigma}_z-\overline{\sigma}_x)^2] \le [k - s(\overline{\sigma}_x + \overline{\sigma}_y + \overline{\sigma}_z)]^2,
\end{equation}
where ($\overline{\sigma}_x$, $\overline{\sigma}_y$, $\overline{\sigma}_z$) are the three average orthogonal shear stresses, $k$ is the internal cohesion and $s$ is a slope constant determined by the angle of friction $\phi$ which was  measured on lunar regolith as $40^\circ$ \citep{Mitchell1974}
\begin{equation}
  s = {2\sin\phi \over \sqrt{3}(3 - \sin\phi)}.
\end{equation}
The ($\overline{\sigma}_x$, $\overline{\sigma}_y$, $\overline{\sigma}_z$) can be calculated by
\begin{equation}
  \overline{\sigma}_x = {(\rho \omega^2 - 2 \pi \rho^2 G A_x) a^2 \over 5},
\end{equation}
\begin{equation}
  \overline{\sigma}_y = {(\rho \omega^2 - 2 \pi \rho^2 G A_y) b^2 \over 5},
\end{equation}
\begin{equation}
  \overline{\sigma}_z = {(-2 \pi \rho^2 G A_z) c^2 \over 5},
\end{equation}
where $\rho$ is the bulk density, $\omega$ is the spin rate, $G$ is the gravitational constant, ($a$, $b$, $c$) are the axes of the asteroid ellipsoidal shape in which $a \ge b \ge c$. Moreover, the ($A_x$, $A_y$, $A_z$) are dimensionless constants that depends on the axial ratios:
\begin{equation}
  A_x = \int_{0}^{\infty} {du \over (u+1)^{3/2}(u+\beta^2){1/2}(u+\alpha^2)^{1/2}},
\end{equation}
\begin{equation}
  A_x = \int_{0}^{\infty} {du \over (u+1)^{1/2}(u+\beta^2){3/2}(u+\alpha^2)^{1/2}},
\end{equation}
\begin{equation}
  A_x = \int_{0}^{\infty} {du \over (u+1)^{1/2}(u+\beta^2){1/2}(u+\alpha^2)^{3/2}},
\end{equation}
where $\alpha = c/a$ and $\beta = b/a$. We assume $a > b = c$, and use $\Delta M = 1.0$ mag\footnote{the amplitude of the lightcurve was not corrected for phase angle effects due to its low-phase-angle (i.e., $<$ 2 degree) observations.} to calculate $a/b = 2.51$ from $10^{0.4 \Delta M}$ for 1999 RE88. Using the average $\rho = 2.72$~g/cm$^3$  for typical S-type asteroids \citep{Demeo2013}, a cohesive strength of $780 \pm 500 Pa$\footnote{The uncertainty of the cohesion includes (a) the derived rotation period, in which we consider the other two solutions beside the best-fit solution as the upper/lower limits (i.e, $P = 1.96 \pm 0.08$ hr), (b) the lightcurve amplitude $\Delta m = 1.0 \pm 0.1$ mag, and (c) the $WISE$/$NEOWISE$ diameter estimation $D = 1.9 \pm 0.3$ km.} would be required to keep the fast-rotating 1999 RE88 intact. Combining with the cohesive strengthes of other known SFRs [i.e., 2001 OE84, $\sim 1500~ Pa$\footnote{The cohesion is calculated here with the parameters in \citet{Pravec2002}}; 2005 UW163, $\sim 200~ Pa$\footnote{The cohesion is calculated here with the parameters in \citet{Chang2014b}}; 2000 GD65, 150 to 450 $Pa$ \citep{Polishook2016}; 1950 DA, 64 $Pa$ \citep{Rozitis2014}], all of them fall within the cohesion range of 100 to 1000 $Pa$ of the lunar regolith \citep{Mitchell1974}. This probably indicates the typical range of internal cohesion for asteroids. When assuming $\rho = 2$~g/cm$^3$ for other 18 SFR candidates, we found that seven candidates with diameter of few kilometers require $> 1000~Pa$ cohesive strengthes and the highest value can be up to 4000 $Pa$ (see last column in Table~\ref{table_sfr}). Therefore, confirming the aforementioned SFR candidates, especially those requiring unusually large cohesive strength, can provide important constraints on the asteroid interior structure.

\subsection{The Spin-Rate Distribution}
\begin{figure}
\plotone{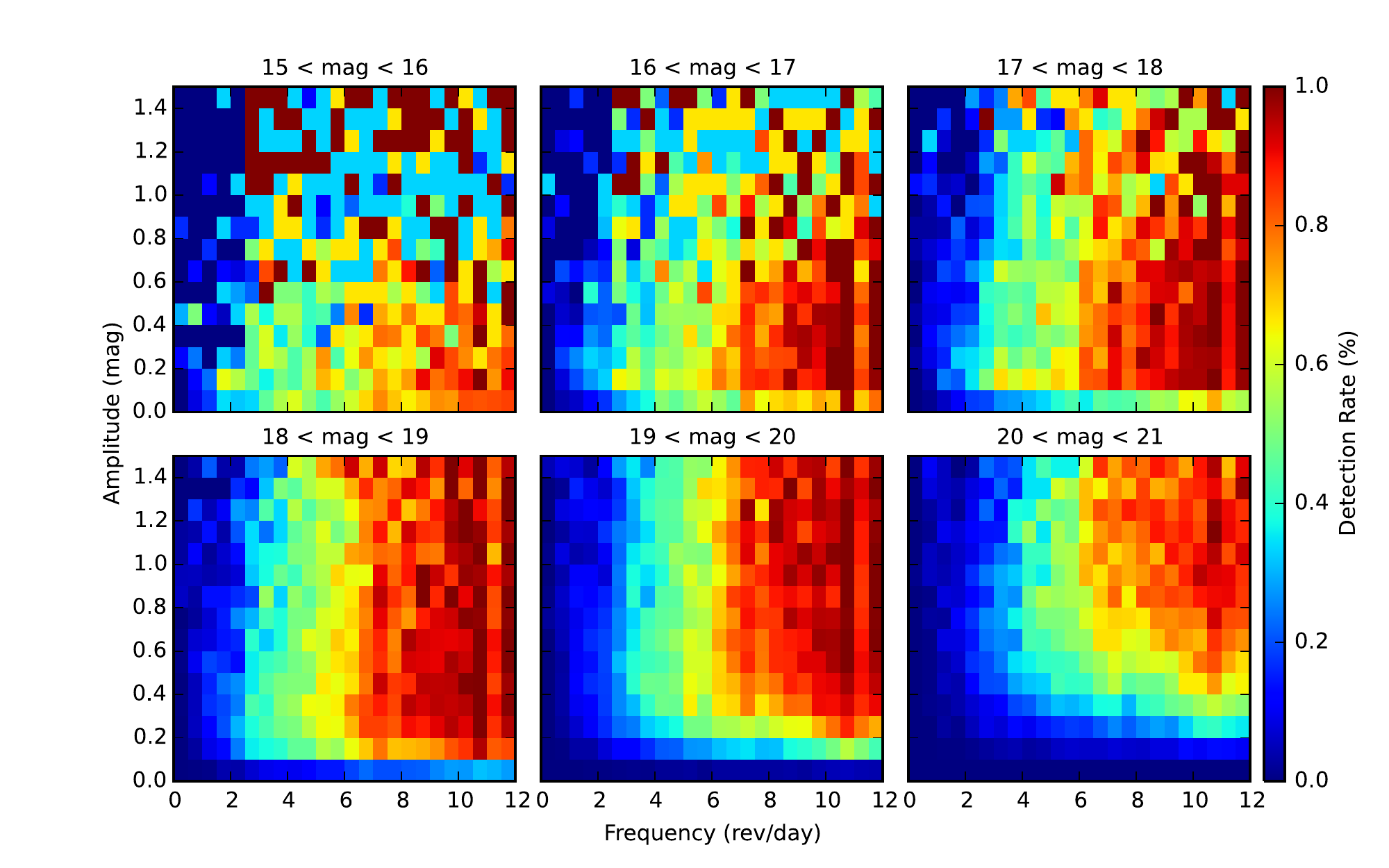}
\caption{Detection rates for asteroid rotation period. The color bar scale on the right shows the percentage
of successful recovery for rotation period of synthetic objects. The apparent magnitude interval is indicated
on the top of each plot.}
\label{debias_map}
\end{figure}

In order to understand any possible observational bias in our survey, we followed the approach of \citet{Chang2015} to carry out a detection efficiency simulation (see \citet{Chang2015} and the references therein). Fig~\ref{debias_map} shows the detection efficiency of spin rate vs. lightcurve amplitude, in which we see that: (a) spin rates of $f \le 3$ rev/day can not be recovered; (b) the detection efficiency of $3 < f < 5$ rev/day is $\lesssim 40\%$, which is much lower than $\sim 90\%$ in \citet{Chang2015}; (c) the spin rates of the asteroids with small lightcurve amplitude (i.e., $\Delta m < 0.1$ mag) are merely to be resolved; and (d) the detection efficiency decreases along with increase of magnitude. The first two situations are due to our two-day observation time-span, which hinders the recovery of relatively long rotation periods. The last two situations can be explained by the photometric uncertainty. When the asteroid's brightness variation is smeared in the photometric uncertainty, the rotation period is not likely to be recovered.

\begin{figure}
\plotone{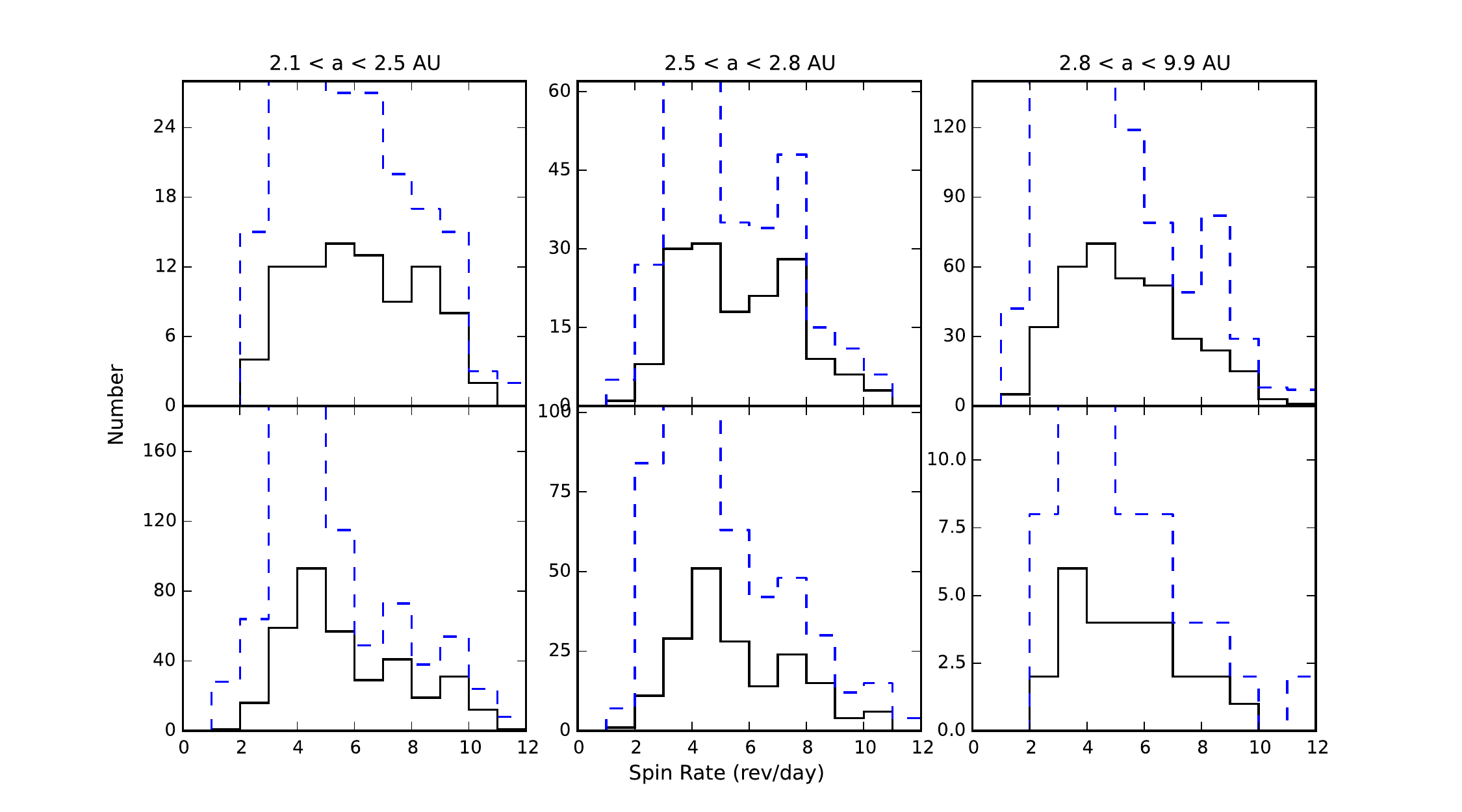}
\caption{The spin-rate distributions for asteroids with diameter of $3 < D < 15$ km (top) and $D <  3$ km (bottom) at inner (left), mid (middle) and outer (right) main-belt. The black line and blue dashed line are the original and de-biased results, respectively.}
\label{spin_rate_comp}
\end{figure}

With this detection efficiency simulation, we generate the de-biased spin-rate distributions and show it along with the original distributions in Fig.~\ref{spin_rate_comp}, in which we separate the distributions according to asteroids' diameters (i.e., $3 < D < 15$ km and $D < 3$ km) and locations in the main belt (i.e., inner: $2.1 < a < 2.5$ AU, mid: $2.5 < a < 2.8$ AU, and outer: $a > 2.8$ AU). Note that the PTF-Ps are not included in the statistics due to their relatively large uncertainties. As expected from the detection efficiency simulation, we see a significant underestimation in the number of $f \le 3$ rev/day and obvious differences between the original and de-biased results for $f \le 5$ rev/day. Although the original distributions looks different from \citet{Chang2015} (i.e., almost no objects in $f \le 2$ rev/day and relatively less objects in $ 3 \le f \le 5$ rev/day in this work), the de-biased results remain consistent in two ways: a) for asteroids of $3 < D < 15$ km, the number in each spin-rate bin decreases along with increase of frequency for $f > 5$ rev/day; and b) for asteroids of $D < 3$ km, a significant number drop at $f = 5$ rev/day (i.e., the number of $f = 6$ rev/day is only half of that in $f = 5$ rev/day).

We aimed to discover large SFRs in this study, and therefore, the observation time-span of each campaign was chosen to be two days in order to obtain sky coverage as large as possible. Although this approach sacrificed the rotation period recovery rate, especially for relatively long periods, the quality of the spin-rate statistics still remains acceptable as a byproduct of our main goal.

\section{Summary}
Five surveys for discovering large SFRs were carried out by using the iPTF. Out of 1029 reliable rotation periods, we found one large SFR, (40511) 1999 RE88 and other 18 candidates. 1999 RE88 is a S-type inner main-belt asteroid with a diameter of $D = 1.9 \pm 0.3$ km, which completes one rotation in $1.96 \pm 0.01$ hr and has a lightcurve amplitude of $\sim 1.0$ mag. This requires an internal cohesion of $\sim 780~ Pa$ for 1999 RE88 to remain intact under such fast rotation in the context of the rubble-pile model. Combining with other known large SFRs, their population seems to be relatively small compared to the entire asteroid population. This indicates that the large SFRs are probably a special group among asteroids.

Although the time span of just two days reduces the rotation period recovery, the spin-rate distributions is in a good agreement with the result of \citet{Chang2015}, which shows number decrease along with increase of spin-rate for asteroids of $3 < D < 15$ km and a significant number drop at $f = 6$ rev/day for asteroids of $D < 3$ km.

\acknowledgments We would like to thank the anonymous referee for his useful suggestions and coments. This work is supported in part by the National Science Council of Taiwan under the grants MOST 104-2112-M-008-014-MY3 and MOST 104-2119-M-008-024, and also by Macau Science and Technology Fund No. 017/2014/A1 of MSAR. This publication makes use of data products from $WISE$, which is a joint project of the University of California, Los Angeles, and the Jet Propulsion Laboratory/California Institute of Technology, funded by the National Aeronautics and Space Administration. This publication also makes use of data products from $NEOWISE$, which is a project of the Jet Propulsion Laboratory/California Institute of Technology, funded by the Planetary Science Division of the National Aeronautics and Space Administration. We gratefully acknowledge the extraordinary services specific to $NEOWISE$ contributed by the International Astronomical Union's Minor Planet Center, operated by the Harvard-Smithsonian Center for Astrophysics, and the Central Bureau for Astronomical Telegrams, operated by Harvard University.

\clearpage

\clearpage

\clearpage

\clearpage

\clearpage

\clearpage

\clearpage
\begin{deluxetable}{lrrcccc}
\tabletypesize{\scriptsize}
\tablecaption{Survey observations in late 2014 and early 2015. \label{obs_log_01}}
\tablewidth{0pt}
\startdata \tableline\tableline
Field ID & RA & Dec. & Oct 29 2014 & Oct 30 2014 & Oct 31 2014 \\
         & ($^{\circ}$) & ($^{\circ}$) & $\Delta$t, N$_\textrm{exp}$ & $\Delta$t, N$_\textrm{exp}$ & $\Delta$t, N$_\textrm{exp}$ \\
\tableline
    3019 &  25.71 &   7.88 &     4.1, 16 &    4.8, 28 &    3.2, 20 \\
    3124 &  25.96 &  10.12 &     4.1, 17 &    4.8, 28 &    3.0, 19 \\
    3125 &  29.42 &  10.12 &     2.7, 11 &    4.8, 28 &    3.2, 20 \\
    3228 &  26.21 &  12.38 &     4.1, 12 &    4.8, 28 &    3.0, 19 \\
    3229 &  29.71 &  12.38 &     4.0, 15 &    4.8, 28 &    3.2, 20 \\
    3332 &  30.00 &  14.62 &     3.8, 13 &    4.8, 28 &    3.2, 20 \\
\tableline
    \\
\tableline\tableline
Field ID & RA & Dec. & Nov 10 2014 & Nov 11 2014 & Nov 13 2014\\
         & ($^{\circ}$) & ($^{\circ}$) & $\Delta$t, N$_\textrm{exp}$ & $\Delta$t, N$_\textrm{exp}$ & $\Delta$t, N$_\textrm{exp}$ \\
\tableline
    3125 &  29.42 &  10.12 &     1.6,  7 &    5.0, 30 &    1.0,  6 \\
    3229 &  29.71 &  12.38 &     1.7,  9 &    5.0, 30 &    2.1,  6 \\
    3230 &  33.20 &  12.38 &     1.8,  9 &    5.0, 30 &    0.3,  3 \\
    3332 &  30.00 &  14.62 &     1.7,  8 &    5.0, 30 &    1.5,  6 \\
    3333 &  33.53 &  14.62 &     1.7,  8 &    5.0, 30 &    1.9,  6 \\
    3435 &  33.86 &  16.88 &     1.7,  8 &    5.0, 30 &    0.3,  3 \\
\tableline
   \\
\tableline\tableline
Field ID & RA & Dec. & Jan 18 2015 & Jan 19 2015 \\
         & ($^{\circ}$) & ($^{\circ}$) & $\Delta$t, N$_\textrm{exp}$ & $\Delta$t, N$_\textrm{exp}$ \\
\tableline
    3559 & 117.00 &  19.12 &    5.7, 34 &    5.9, 36 \\
    3560 & 120.60 &  19.12 &    5.7, 34 &    5.9, 36 \\
    3561 & 124.20 &  19.12 &    5.7, 34 &    5.9, 36 \\
    3562 & 127.80 &  19.12 &    5.7, 33 &    5.9, 36 \\
    3563 & 131.40 &  19.12 &    5.7, 34 &    5.9, 36 \\
    3564 & 135.00 &  19.12 &    5.7, 35 &    5.9, 36 \\
\tableline
   \\
\tableline\tableline
Field ID & RA & Dec. & Jan 20 2015 & Jan 21 2015 \\
         & ($^{\circ}$) & ($^{\circ}$) & $\Delta$t, N$_\textrm{exp}$ & $\Delta$t, N$_\textrm{exp}$ \\
\tableline
    3461 & 126.53 &  16.88 &    6.1, 35 &    5.5, 34 \\
    3462 & 130.10 &  16.88 &    6.2, 35 &    5.5, 34 \\
    3463 & 133.66 &  16.88 &    6.2, 35 &    5.5, 34 \\
    3464 & 137.23 &  16.88 &    6.2, 36 &    5.5, 34 \\
    3465 & 140.79 &  16.88 &    6.2, 36 &    5.5, 34 \\
    3466 & 144.36 &  16.88 &    6.2, 36 &    5.5, 34 \\
\tableline
   \\
\tableline\tableline
Field ID & RA & Dec. & Feb 25 2015 & Feb 26 2015 \\
         & ($^{\circ}$) & ($^{\circ}$) & $\Delta$t, N$_\textrm{exp}$ & $\Delta$t, N$_\textrm{exp}$ \\
\tableline
    3159 & 147.12 &  10.12 &    4.6, 19 &    4.5, 28 \\
    3160 & 150.58 &  10.12 &    4.6, 20 &    4.5, 28 \\
    3161 & 154.04 &  10.12 &    4.6, 21 &    4.5, 28 \\
    3162 & 157.50 &  10.12 &    4.5, 18 &    4.5, 28 \\
    3163 & 160.96 &  10.12 &    4.5, 21 &    4.6, 28 \\
    3164 & 164.42 &  10.12 &    4.5, 22 &    4.6, 28 \\
\tableline

\enddata
\tablecomments{$\Delta$t is the time duration spanned by each observing set in hours and
N$_\textrm{exp}$ is the total number of exposures for each night and field.}
\end{deluxetable}

\begin{deluxetable}{llrrrrrrrrrrlrrrrl}
\tabletypesize{\scriptsize} \setlength{\tabcolsep}{0.02in} \tablecaption{Synodic rotation periods of
PTF-U2s. \label{table_p}} \tablewidth{0pt} \tablehead{ \colhead{Obj ID} & \colhead{Designation} &
\colhead{$a$} & \colhead{$e$} & \colhead{$i$} & \colhead{$\Omega$} & \colhead{$\omega$} & \colhead{D}
& \colhead{$\triangle$} & \colhead{$r$} & \colhead{$\alpha$} & \colhead{$H_R$} & \colhead{n} &
\colhead{m} & \colhead{$PTF_R$} & \colhead{Period (hr)} & \colhead{$\triangle m$} & \colhead{U}}
\startdata
00435$^{*}$       & (435) Ella                   &  2.45 &  0.16 &  1.82 &  23.2 & 333.6 &  23.3       &   2.69 &   1.74 &  7.00 & 10.18$\pm$0.13   &  2 &  70 & 14.04$\pm$0.00   &   4.64$\pm$0.02  &  0.36 &  3 \\
00492$^{*}$       & (492) Gismonda               &  3.11 &  0.18 &  1.62 &  46.2 & 296.5 &  59.9$^w$   &   3.66 &   2.68 &  0.90 &  9.68$\pm$0.07   &  2 &  47 & 14.88$\pm$0.00   &   6.44$\pm$0.04  &  0.14 &  2 \\
00996$^{*}$       & (996) Hilaritas              &  3.09 &  0.14 &  0.66 & 347.4 & 147.2 &  30.9$^w$   &   2.66 &   1.68 &  3.33 & 10.87$\pm$0.21   &  7 &  68 & 14.25$\pm$0.00   &   9.70$\pm$0.10  &  0.54 &  2 \\
01223$^{*}$       & (1223) Neckar                &  2.87 &  0.06 &  2.54 &  40.8 &  14.4 &  25.7$^w$   &   2.71 &   1.73 &  4.28 & 10.16$\pm$0.10   &  3 &  42 & 13.89$\pm$0.00   &   7.80$\pm$0.06  &  0.21 &  2 \\
01635$^{*}$       & (1635) Bohrmann              &  2.85 &  0.06 &  1.82 & 184.3 & 135.4 &  17.5$^w$   &   3.01 &   2.03 &  1.89 & 10.79$\pm$0.10   &  2 &  69 & 14.96$\pm$0.00   &   5.85$\pm$0.04  &  0.34 &  3 \\
01709$^{*}$       & (1709) Ukraina               &  2.38 &  0.21 &  7.56 & 300.1 &  42.5 &   8.3       &   2.73 &   1.75 &  2.69 & 12.42$\pm$0.16   &  2 &  68 & 16.01$\pm$0.01   &   7.33$\pm$0.06  &  0.53 &  3 \\
01850             & (1850) Kohoutek              &  2.25 &  0.13 &  4.05 &  68.9 & 190.6 &   7.6$^w$   &   2.42 &   1.43 &  3.92 & 12.81$\pm$0.09   &  2 &  41 & 15.75$\pm$0.00   &   3.68$\pm$0.01  &  0.31 &  3 \\
01881$^{*}$       & (1881) Shao                  &  3.16 &  0.11 &  9.86 & 218.1 &  67.7 &  25.4$^w$   &   3.24 &   2.25 &  1.01 & 11.19$\pm$0.04   &  2 &  40 & 15.61$\pm$0.00   &   5.61$\pm$0.07  &  0.11 &  2 \\
01983$^{*}$       & (1983) Bok                   &  2.62 &  0.10 &  9.41 &  23.6 & 345.8 &  11.2       &   2.39 &   1.40 &  2.46 & 12.77$\pm$0.15   &  3 &  57 & 15.64$\pm$0.00   &   8.73$\pm$0.08  &  0.40 &  3 \\
02114$^{*}$       & (2114) Wallenquist           &  3.20 &  0.14 &  0.56 &   1.6 & 216.8 &  23.0$^w$   &   3.13 &   2.16 &  3.79 & 11.60$\pm$0.10   &  3 &  68 & 16.07$\pm$0.00   &   5.52$\pm$0.03  &  0.28 &  3 \\
02276$^{*}$       & (2276) Warck                 &  2.37 &  0.17 &  2.47 & 215.4 &  55.2 &  15.1$^w$   &   2.59 &   1.61 &  5.11 & 12.81$\pm$0.09   &  3 &  41 & 16.37$\pm$0.01   &   4.42$\pm$0.04  &  0.24 &  2 \\
&&&&&&&&....&&&&&&&&&\\
\enddata
\tablecomments{Columns: asteroid's designations, semi-major axis ($a$, AU), eccentricity ($e$,
degree), inclination ($i$, degree), longitude of ascending node ($\Omega$, degree), argument of
periapsis ($\omega$, degree), diameter (D, km), heliocentric distance ($\triangle$, AU), geodesic
distance ($r$, AU), phase angle ($\alpha$, degree), absolute magnitude ($H$, mag), number of nights
(n), number of images (m), $PTF_R$ magnitude, derived rotation period (hours), lightcurve amplitude
(mag) and rotation period quality code (U).  The full table is available in the eletronic version.}
\tablenotetext{*}{Asteroid available in the LCDB.} \tablenotetext{b}{Lightcurves with large amplitudes
and deep V-shape minima.} \tablenotetext{w}{$WISE$/$NEOWISE$ diameter.} \tablenotetext{e}{Estimated
diameter from Eq~\ref{dia_eq}.}
\end{deluxetable}

\begin{deluxetable}{llrrrrrrrrrrlrrrrl}
\tabletypesize{\scriptsize} \setlength{\tabcolsep}{0.02in} \tablecaption{Asteroids with partial phase
coverage. \label{table_p_part}} \tablewidth{0pt} \tablehead{ \colhead{Obj ID} & \colhead{Designation}
& \colhead{$a$} & \colhead{$e$} & \colhead{$i$} & \colhead{$\Omega$} & \colhead{$\omega$} &
\colhead{D} & \colhead{$\triangle$} & \colhead{$r$} & \colhead{$\alpha$} & \colhead{$H_R$} &
\colhead{n} & \colhead{m} & \colhead{$PTF_R$} & \colhead{Period (hr)} & \colhead{$\triangle m$} &
\colhead{U}} \startdata
00870$^{*}$       & (870) Manto                  &  2.32 &  0.26 &  6.19 & 120.8 & 196.9 &  13.0       &   2.91 &   1.93 &  3.29 & 11.45$\pm$0.12   &  4 &  77 & 15.34$\pm$0.00   &  30.00$\pm$4.05  &  0.08 &  2 \\
01091             & (1091) Spiraea               &  3.42 &  0.06 &  1.16 &  80.7 &  10.0 &  40.3$^w$   &   3.28 &   2.32 &  4.74 & 10.70$\pm$0.08   &  2 &  35 & 15.48$\pm$0.00   &   7.01$\pm$0.43  &  0.03 &  2 \\
01142$^{*}$       & (1142) Aetolia               &  3.18 &  0.08 &  2.11 & 139.3 &  96.2 &  24.4$^w$   &   3.11 &   2.12 &  0.40 &  9.95$\pm$0.07   &  2 &  48 & 14.05$\pm$0.00   &   7.68$\pm$0.12  &  0.15 &  2 \\
01782$^{*}$       & (1782) Schneller             &  3.11 &  0.16 &  1.54 & 157.4 & 107.2 &  21.9$^w$   &   3.39 &   2.43 &  4.77 & 11.65$\pm$0.20   &  3 &  43 & 16.59$\pm$0.01   &   5.93$\pm$0.07  &  0.71 &  2 \\
02142             & (2142) Landau                &  3.16 &  0.12 &  0.66 & 155.5 &  34.2 &  20.1$^w$   &   3.51 &   2.52 &  2.86 & 11.84$\pm$0.08   &  3 &  55 & 16.86$\pm$0.01   &   9.70$\pm$0.20  &  0.28 &  2 \\
02164             & (2164) Lyalya                &  3.19 &  0.13 &  2.63 & 115.7 & 196.7 &  20.0$^w$   &   3.55 &   2.56 &  1.10 & 11.48$\pm$0.08   &  2 &  48 & 16.58$\pm$0.01   &  11.57$\pm$0.14  &  0.18 &  2 \\
02259$^{*}$       & (2259) Sofievka              &  2.29 &  0.19 &  4.68 & 280.3 &  12.1 &   8.2       &   2.71 &   1.73 &  2.04 & 12.44$\pm$0.06   &  2 &  69 & 16.00$\pm$0.00   &   9.80$\pm$0.31  &  0.10 &  2 \\
02371             & (2371) Dimitrov              &  2.44 &  0.01 &  1.78 & 235.0 & 282.6 &   7.9       &   2.41 &   1.43 &  4.76 & 12.51$\pm$0.06   &  2 &  47 & 15.45$\pm$0.00   &   6.40$\pm$0.13  &  0.06 &  2 \\
02388             & (2388) Gase                  &  2.45 &  0.18 &  2.22 & 324.6 & 253.1 &   6.3       &   2.35 &   1.39 &  6.57 & 13.02$\pm$0.16   &  2 &  70 & 16.08$\pm$0.00   &  13.15$\pm$0.35  &  0.48 &  2 \\
02415$^{*}$       & (2415) Ganesa                &  2.66 &  0.04 &  2.37 &  89.8 & 209.1 &  15.5$^w$   &   2.73 &   1.74 &  2.41 & 12.07$\pm$0.05   &  1 &  20 & 15.71$\pm$0.00   &   2.67$\pm$0.13  &  0.16 &  2 \\
02540             & (2540) Blok                  &  2.20 &  0.05 &  1.27 & 183.5 & 261.3 &   6.1$^w$   &   2.11 &   1.13 &  3.65 & 13.14$\pm$0.02   &  2 &  50 & 15.36$\pm$0.00   &   9.60$\pm$3.67  &  0.07 &  2 \\
02645$^{*}$       & (2645) Daphne Plane          &  2.39 &  0.11 & 13.79 & 349.8 &  79.5 &  10.4       &   2.37 &   1.38 &  3.06 & 11.92$\pm$0.09   &  2 &  45 & 14.78$\pm$0.00   &   7.11$\pm$0.05  &  0.31 &  2 \\
&&&&&&&&....&&&&&&&&&\\
\enddata
\tablecomments{The amplitudes of the objects with partial lightcurve coverage and lightcurves with a
single minimum should be treated as lower limits. Also, see note and footnotes associated with
Table~\ref{table_p} for nomenclature and explanation.}
\end{deluxetable}

\begin{deluxetable}{llrrrrrrrrrrlrrrrlr}
\tabletypesize{\scriptsize} \setlength{\tabcolsep}{0.02in} \tablecaption{The SFR (40511) 1999 RE88 and other 18 candidates.
\label{table_sfr}} \tablewidth{0pt} \tablehead{ \colhead{Obj ID} & \colhead{Designation}
& \colhead{$a$} & \colhead{$e$} & \colhead{$i$} & \colhead{$\Omega$} & \colhead{$\omega$} &
\colhead{D} & \colhead{$\triangle$} & \colhead{$r$} & \colhead{$\alpha$} & \colhead{$H_R$} &
\colhead{n} & \colhead{m} & \colhead{$PTF_R$} & \colhead{Period (hr)} & \colhead{$\triangle m$} &
\colhead{U} & \colhead{$k$}} \startdata
40511             & (40511) 1999 RE88            &  2.38 &  0.17 &  2.04 & 341.6 & 279.8 &   1.9$^w$   &   2.61 &   1.62 &  1.93 & 16.36$\pm$0.30   &  3 &  54 & 19.70$\pm$0.08   &   1.96$\pm$0.01  &  1.04 &  2 &    670  \\
A0351             & (100351) 1995 SU88           &  2.42 &  0.13 &  0.64 & 356.5 & 199.4 &   1.0       &   2.71 &   1.72 &  1.37 & 17.05$\pm$0.20   &  3 &  39 & 20.56$\pm$0.14   &   1.99$\pm$0.01  &  1.00 &  2 &    170  \\
E4977             & (144977) 2005 EC127          &  2.21 &  0.17 &  4.75 & 336.9 & 312.8 &   0.9       &   2.45 &   1.46 &  1.58 & 17.27$\pm$0.19   &  2 &  43 & 20.13$\pm$0.12   &   1.64$\pm$0.01  &  0.72 &  2 &    150  \\
F2066             & (152066) 2004 PT108          &  2.56 &  0.20 &  2.28 & 335.7 & 287.2 &   2.2       &   2.82 &   1.87 &  7.01 & 16.33$\pm$0.25   &  2 &  37 & 20.42$\pm$0.16   &   1.42$\pm$0.01  &  0.93 &  2 &   1740  \\
G8089             & (168089) 2006 DM84           &  2.23 &  0.07 &  1.38 & 230.2 &  27.7 &   0.9       &   2.34 &   1.35 &  1.07 & 17.13$\pm$0.19   &  1 &  24 & 19.79$\pm$0.09   &   1.39$\pm$0.06  &  0.68 &  2 &    220  \\
J1530             & (191530) 2003 UX197          &  3.09 &  0.11 &  5.36 &  49.4 & 193.2 &   5.3       &   3.39 &   2.41 &  3.72 & 15.12$\pm$0.14   &  3 &  36 & 20.01$\pm$0.13   &   1.78$\pm$0.02  &  0.50 &  2 &   2950  \\
K02QC9J           & 2002 QJ129                   &  3.03 &  0.14 & 10.04 & 123.8 & 274.2 &   3.5       &   2.92 &   1.94 &  1.05 & 16.05$\pm$0.16   &  3 &  64 & 19.95$\pm$0.11   &   1.71$\pm$0.02  &  0.51 &  2 &   1480  \\
K05S16S           &                              &  2.70 &  0.09 &  1.22 & 222.4 & 181.1 &   1.3       &   2.46 &   1.48 &  3.23 & 17.50$\pm$0.18   &  3 &  45 & 20.65$\pm$0.13   &   1.89$\pm$0.01  &  0.71 &  2 &    200  \\
K08D80Z           &                              &  2.37 &  0.13 &  1.02 & 185.3 &  12.9 &   0.6       &   2.26 &   1.28 &  1.29 & 18.00$\pm$0.16   &  2 &  45 & 20.46$\pm$0.12   &   1.44$\pm$0.01  &  0.63 &  2 &     80  \\
K08UU6L           &                              &  3.07 &  0.16 &  4.01 & 278.7 & 141.4 &   2.4       &   2.79 &   1.81 &  1.67 & 16.89$\pm$0.15   &  2 &  52 & 20.60$\pm$0.13   &   1.59$\pm$0.01  &  0.59 &  2 &    960  \\
K15C40P           &                              &  3.07 &  0.03 &  9.52 & 149.9 &  82.7 &   3.8       &   3.04 &   2.05 &  0.82 & 15.89$\pm$0.18   &  2 &  43 & 19.97$\pm$0.12   &   1.27$\pm$0.01  &  0.65 &  2 &   4650  \\
L7411             & (217411) 2005 LD50           &  2.45 &  0.25 & 10.91 & 297.5 &  16.5 &   1.5       &   3.05 &   2.07 &  2.28 & 16.20$\pm$0.19   &  2 &  60 & 20.41$\pm$0.13   &   1.91$\pm$0.02  &  0.73 &  2 &    270  \\
L8602             & (218602) 2005 NE69           &  2.41 &  0.12 &  1.13 & 253.2 &  14.7 &   1.0       &   2.65 &   1.66 &  1.47 & 16.97$\pm$0.22   &  3 &  55 & 20.38$\pm$0.14   &   1.55$\pm$0.01  &  0.86 &  2 &    260  \\
P3384             & (253384) 2003 KQ3            &  2.16 &  0.19 &  4.97 & 297.6 & 343.9 &   0.8       &   2.53 &   1.55 &  2.68 & 17.44$\pm$0.19   &  3 &  51 & 20.68$\pm$0.14   &   1.47$\pm$0.01  &  0.71 &  2 &    160  \\
P5828             & (255828) 2006 SC86           &  2.40 &  0.14 &  1.16 & 193.8 & 269.7 &   0.6       &   2.07 &   1.09 &  1.90 & 18.02$\pm$0.16   &  3 &  64 & 19.88$\pm$0.09   &   1.73$\pm$0.02  &  0.47 &  2 &     40  \\
Q8611             & (268611) 2006 CY30           &  2.73 &  0.03 &  6.34 & 169.6 & 181.7 &   2.4       &   2.81 &   1.84 &  4.56 & 16.14$\pm$0.15   &  2 &  37 & 20.10$\pm$0.11   &   1.46$\pm$0.01  &  0.61 &  2 &   1240  \\
W6242             & (326242) 2012 DS21           &  2.55 &  0.08 &  0.86 & 143.1 &  22.6 &   1.6       &   2.68 &   1.69 &  2.01 & 17.02$\pm$0.21   &  3 &  52 & 20.45$\pm$0.13   &   1.04$\pm$0.00  &  0.75 &  2 &   1480  \\
X2984             & (332984) 2011 FG67           &  3.19 &  0.04 & 13.17 &  32.0 &  75.8 &   4.5       &   3.14 &   2.17 &  4.29 & 15.47$\pm$0.16   &  3 &  40 & 19.97$\pm$0.13   &   1.59$\pm$0.01  &  0.64 &  2 &   3630  \\
b2963             & (372963) 2011 BY111          &  2.66 &  0.11 &  0.63 & 167.7 & 249.9 &   1.1       &   2.48 &   1.50 &  0.72 & 17.81$\pm$0.18   &  2 &  42 & 20.64$\pm$0.14   &   1.07$\pm$0.00  &  0.66 &  2 &    590  \\
\enddata
\tablecomments{The cohesion $k$ is calculated assuming bulk density $\rho = 2$~g/cm$^3$, except for (40511) 1999 RE88. Also, see note and footnotes associated with Table~\ref{table_p} for nomenclature and explanation.}
\end{deluxetable}


\begin{thebibliography}{}
\bibitem[Bowell et al.(1989)]{Bowell1989} Bowell, E., Hapke, B., Domingue, D., et al.\ 1989, Asteroids II, 524
\bibitem[Chang et al.(2014a)]{Chang2014a} Chang, C.-K., Ip, W.-H., Lin, H.-W., et al.\ 2014a, \apj, 788, 17
\bibitem[Chang et al.(2014b)]{Chang2014b} Chang, C.-K., Waszczak, A., Lin, H.-W., et al.\ 2014b, \apjl, 791, LL35
\bibitem[Chang et al.(2015)]{Chang2015} Chang, C.-K., Ip, W.-H., Lin, H.-W., et al.\ 2015, \apjs, 219, 27
\bibitem[Chang et al.(2016)]{Chang2016} Chang, C.-K., Lin, H.-W., \& Ip, W.-H.\ 2016, \apj, 816, 71
\bibitem[DeMeo \& Carry(2013)]{Demeo2013} DeMeo, F.~E., \& Carry, B.\ 2013, \icarus, 226, 723
\bibitem[Dermawan et al.(2011)]{Dermawan2011} Dermawan, B., Nakamura, T., \& Yoshida, F.\ 2011, \pasj, 63, 555
\bibitem[Grav et al.(2011)]{Grav2011} Grav, T., Mainzer, A.~K., Bauer, J., et al.\ 2011, \apj, 742, 40
\bibitem[Grillmair et al.(2010)]{Grillmair2010} Grillmair, C.~J., Laher, R., Surace, J., et al.\ 2010, Astronomical Data Analysis Software and Systems XIX, 434, 28
\bibitem[Harris et al.(1989)]{Harris1989} Harris, A.~W., Young, J.~W., Bowell, E., et al.\ 1989, \icarus, 77, 171
\bibitem[Harris(1996)]{Harris1996} Harris, A.~W.\ 1996, Lunar and Planetary Institute Science Conference Abstracts, 27, 493
\bibitem[Harris et al.(2014)]{Harris2014} Harris, A.~W., Pravec, P., Gal{\'a}d, A., et al.\ 2014, \icarus, 235, 55
\bibitem[Holsapple(2007)]{Holsapple2007} Holsapple, K.~A.\ 2007, \icarus, 187, 500
\bibitem[Jewitt, Ishiguro \& Agarwal (2013)]{Jweitt2013} Jewitt, D., Ishiguro, M., \& Agarwal,  J.\ 2013, \apjl, 764, L5
\bibitem[Laher et al.(2014)]{Laher2014} Laher, R.~R., Surace, J., Grillmair, C.~J., et al.\ 2014, \pasp, 126, 674
\bibitem[Law et al.(2009)]{Law2009} Law, N.~M., Kulkarni, S.~R., Dekany, R.~G., et al.\ 2009, \pasp, 121, 1395
\bibitem[Law et al.(2010)]{Law2010} Law, N.~M., Dekany, R.~G., Rahmer, G., et al.\ 2010, \procspie, 7735
\bibitem[Mainzer et al.(2011)]{Mainzer2011} Mainzer, A., Grav, T., Bauer, J., et al.\ 2011, \apj, 743, 156
\bibitem[Masiero et al.(2009)]{Masiero2009} Masiero, J., Jedicke, R., {\v D}urech, J., et al.\ 2009, \icarus, 204, 145
\bibitem[Masiero et al.(2011)]{Masiero2011} Masiero, J.~R., Mainzer, A.~K., Grav, T., et al.\ 2011, \apj, 741, 68
\bibitem[Mazzone (2012)]{Mazzone2012}Mazzone, F.\ 2012, http://www.astrosurf.com/salvador/Fotometria.html
\bibitem[Mitchell (1974)]{Mitchell1974}Mitchell, J. K., Houston, W. N., Carrier, W. D. \& Costes, N. C.\ 1974, Apollo soil mechanics experiment S-200 final report. Space Sciences Laboratory Series 15, 7285 (Univ. California, Berkeley, 1974)
\bibitem[Ofek et al.(2012a)]{Ofek2012a} Ofek, E.~O., Laher, R., Law, N., et al.\ 2012a, \pasp, 124, 62
\bibitem[Ofek et al.(2012b)]{Ofek2012b} Ofek, E.~O., Laher, R., Surace, J., et al.\ 2012b, \pasp, 124, 854
\bibitem[Polishook and Brosch(2009)]{Polishook2009} Polishook, D., Brosch, N.\ 2009, \icarus, 199, 319
\bibitem[Polishook et al.(2012)]{Polishook2012} Polishook, D., Ofek, E.~O., Waszczak, A., et al.\ 2012, \mnras, 421, 2094
\bibitem[Polishook et al.(2016)]{Polishook2016} Polishook, D., Moskovitz, N., Binzel, R.~P., et al.\ 2016, \icarus, 267, 243
\bibitem[Pravec \& Harris(2000)]{Pravec2000} Pravec, P., \& Harris, A.~W.\ 2000, \icarus, 148, 12
\bibitem[Pravec et al.(2002)]{Pravec2002} Pravec, P., Ku{\v s}nir{\'a}k, P., {\v S}arounov{\'a}, L., et al.\ 2002, Asteroids, Comets, and Meteors: ACM 2002, 500, 743
\bibitem[Pravec et al.(2008)]{Pravec2008} Pravec, P., Harris, A.~W., Vokrouhlick{\'y}, D., et al.\ 2008, \icarus, 197, 497
\bibitem[Rau et al.(2009)]{Rau2009} Rau, A., Kulkarni, S.~R., Law, N.~M., et al.\ 2009, \pasp, 121, 1334
\bibitem[Rozitis et al.(2014)]{Rozitis2014} Rozitis, B., Maclennan, E., \& Emery, J.~P.\ 2014, \nat, 512, 174
\bibitem[Rubincam(2000)]{Rubincam2000} Rubincam, D.~P.\ 2000, \icarus, 148, 2
\bibitem[S{\'a}nchez \& Scheeres(2012)]{Sanchez2012} S{\'a}nchez, D.~P., \& Scheeres, D.~J.\ 2012, \icarus, 218, 876
\bibitem[Tedesco, Cellino \& Zappal{\'a} (2005)]{Tedesco2005} Tedesco, E.~F., Cellino, A., \& Zappal{\'a}, V.\ 2005, \aj, 129, 2869
\bibitem[Warner, Harris \& Pravec (2009)]{Warner2009} Warner, B.~D., Harris, A.~W., \& Pravec, P.\ 2009, \icarus, 202, 134
\bibitem[Waszczak et al.(2015)]{Waszczak2015} Waszczak, A., Chang, C.-K., Ofek, E.~O., et al.\ 2015, \aj, 150, 75
\bibitem[York et al.(2000)]{York2000} York, D.~G., Adelman, J., Anderson, J.~E., Jr., et al.\ 2000, \aj, 120, 1579

\end{thebibliography}
\end{document}